\documentclass[12pt,preprint]{aastex}
\usepackage{epstopdf}
\usepackage{graphicx}
\usepackage{amsmath,amsfonts}
\shortauthors{GWINN ET AL.}
\shorttitle{SIZE OF PULSAR EMISSION REGION}
\begin{document}
\input epsf

\title{Size of the Vela Pulsar's Emission Region at 18 cm Wavelength}
\author{C. R. Gwinn, M. D. Johnson}
\affil{Department of Physics, University of California, Santa Barbara, California 93106}
\email{cgwinn@physics.ucsb.edu} 

\author{J. E. Reynolds, D. L. Jauncey\altaffilmark{1}, and A. K. Tzioumis}
\affil{Australia Telescope National Facility, CSIRO, P.O. Box 76, Epping, NSW 1710, Australia}

\author{H.Hirabayashi, H. Kobayashi\altaffilmark{2}, Y. Murata, and P. G. Edwards\altaffilmark{3}}
\affil{Institute of Space and Astronautical Science, Japan Aerospace Exploration Agency, 3-1-1 Yoshinodai, Chuou-ku Sagamihara, Kanagawa 252-5210, Japan}

\author{S. Dougherty, B. Carlson, and D. del Rizzo}
\affil{National Research Council of Canada, Herzberg Institute for Astrophysics, Dominion Radio Astrophysical Observatory, PO Box 248, Penticton, BC, V2A 6J9, Canada}

\author{J. F. H. Quick and C. S. Flanagan}
\affil{Hartebeesthoek Radio Astronomy Observatory, P.O. Box 443, Krugersdorp 1740, South Africa}

\author{P. M. McCulloch}
\affil{School of Mathematics and Physics, University of Tasmania, Private Bag 37, Hobart, TAS 7001, Australia}
\altaffiltext{1}{Mount Stromlo Observatory, Cotter Road Weston ACT 2611 Australia}
\altaffiltext{2}{Present address: National Astronomical Observatory of Japan, 2-21-1 Osawa, Mitaka, Tokyo, 181-8588, Japan}
\altaffiltext{3}{Present address: Australia Telescope National Facility, CSIRO, P.O. Box 76, Epping, NSW 1710, Australia}

\vskip 1 truein
\begin{abstract}
We present measurements of the linear diameter of the emission region of the Vela pulsar at observing wavelength $\lambda=18\ {\rm cm}$.  We infer the diameter as a function of pulse phase from the distribution of visibility on the Mopra-Tidbinbilla baseline. As we demonstrate, in the presence of strong scintillation, finite size of the emission region produces a characteristic W-shaped signature in the projection of the visibility distribution onto the real axis. This modification involves heightened probability density near the mean amplitude, decreased probability to either side, and a return to the zero-size distribution beyond. We observe this signature with high statistical significance, as compared with the best-fitting zero-size model, in many regions of pulse phase. We find that the equivalent full width at half maximum of the pulsar's emission region decreases from more than 400\ km early in the pulse to near zero at the peak of the pulse, and then increases again to approximately 800\ km near the trailing edge. We discuss possible systematic effects, and compare our work with previous results.
\end{abstract}


\section{INTRODUCTION}

Pulsars emit strong radio emission from compact regions.
The enormous magnetic fields and rapid rotation of neutron stars easily accelerate electrons and positrons to high energy,
but the means by which a small fraction of that energy is transformed to radio emission remains
poorly understood.
Pulsar emission regions are small, but interstellar scattering of radio waves provides an astronomical-unit-scale 
lens with the nanoarcecond resolution sufficient to resolve them spatially.
However, the lens is highly corrupt, and unraveling source structure involves application of statistical models to large volumes of high-quality data. 
These models must include accurate descriptions of the effects of scattering and of noise.

In this paper, we describe observations and data analysis to fit a simple model for a spatially-extended emission region to the
interferometric visibility statistics of the Vela pulsar.
We focus on description of the models, the data, and comparisons of the two.  
In this introductory section, we briefly describe, as background, the features of interstellar scattering important for our technique
and the basic features of pulsar physics involved in the interpretation.

\subsection{Pulsar Emission Physics}

Although pulsars have been observed for more than 40 years, the 
process by which a rapidly-rotating, magnetized neutron star converts a small fraction
of its rotational energy to radio waves remains unclear.
The rapid rotation would produce $\vec v\times \vec B$
forces and induced electric fields ample to tear electrons from the
surface of the neutron star, were those forces not cancelled by an induced
co-rotating charge distribution \citep{gol69}.
The magnetic field of the neutron star away from its surface is nearly dipole, modified by inertia of the corotating particles
and relativistic effects \citep{spi06}.
``Open'' field lines pass through the light cylinder,
the surface where the co-rotation speed is that of light.
These field lines carry highly-relativistic charged particles away from the star,
forming a powerful wind.  
A small fraction of the energy of this wind is apparently converted to radio emission,
observed as pulses because of stellar rotation.

The boundaries of the set of open field lines form promising places for the origin of pulsar emission.
Above the ``polar cap'' at the base of the open field lines, the current may be insufficient to replace the outflowing charge,
so that a gap may form, with strong electric field parallel to the nearly-vertical magnetic field \citep{rud75,aro79}.
Similarly, a gap may form between the last closed field lines, which nearly graze the light cylinder, and the open field lines. 
Proposed locations include a ``slot gap" extending to high altitude from the polar cap \citep{mus04} and 
charge-free ``outer gaps" in the outer magnetosphere \citep{che86,che00}. 
Within these gaps, electrons and positrons accelerate
to TeV energies, accompanied by pair creation.
These particles can emit X-rays and gamma-rays via curvature, synchrotron, and inverse Compton emission.
Force-free simulations of pulsar magnetospheres 
cannot include gaps, but interestingly predict strong currents in the same locations,
which might produce particle acceleration via plasma instabilities 
\citep{spi06,Gru07}.
These particle-acceleration mechanisms may provide the power source for radio emission.

The radio emission is coherent: the observed $\sim 10^{26}$\ K brightness temperature exceeds that possible for individual
electrons \citep{Man77}.  
The nanosecond 
variability of giant pulses indicates that the emission originates in structures $\sim 1$ m across \citep{Han03}.
These structures are likely distributed over scales comparable to those
of the particle-acceleration zones: larger than the polar cap, $\sim 1$ km,
but smaller than the diameter of the light-cylinder.
In this work, we seek to determine the lateral dimension of this region of emission.

Emission could arise directly from particles traveling along the field lines,
in which case polarization and temporal variations would directly reflect conditions at the source.
Alternatively, emission may be reprocessed before it leaves the pulsar's magnetosphere,
as is suggested by observations and theoretical models \citep{lyu00,Jes10}.
Radio emission could also arise
via production of plasma waves from plasma instabilities on open field lines.
These plasma waves would propagate nearly along field lines, 
and then be converted to radio waves where the local plasma frequency falls below the observed frequency \citep{Bar86}.
Our measurements can contribute to this picture by describing the lateral scale of the emission region.

\subsection{Interstellar Scattering and Scintillation}

Radio waves emitted from a pulsar encounter variations in refractive
index in the interstellar medium, from variations in electron density.
Waves travel onward with ``crinkled'' phase fronts and arrive at an observer from along a number of paths
to form a diffraction pattern.
For the interstellar medium at decimeter or longer wavelengths, 
the differences in path lengths are many wavelengths,
and many paths contribute to the diffraction pattern at the observer
\citep{coh74}. 

The diffraction pattern at the observer is the convolution of an image of the source
with the pattern for a point source \citep{cor89,goo96}.
This results from the fact that, for small deflections,
the Kirchoff integrals that relate the electric field at the observer to those at the source
become Fourier transforms, with the effects of the scattering medium inserted multiplicatively at the screen \citep{Gwi98}.
Geometrical factors, depending on the position of the screen, relate the 
original source and corrupt image by a magnification factor
$M=D/R$,
where $D$ is the distance from observer to scatterer, and $R$ is the distance from scatterer to source.

Because path differences are thousands of wavelengths,
paths that reinforce at one observing frequency and position in the observer plane may cancel at nearby frequencies or positions.
For an observed angular extent $\theta_{\rm ISS}$ of the scattering disk and observing wavelength $\lambda$, 
the scale of the diffraction pattern at the observer is $S_{\rm ISS} = \lambda/\theta_{\rm ISS}$,
equal to the linear resolution of the scattering disk viewed as a lens.
The source shows scintillations, or intensity variations, 
with timescale $\Delta t_{\rm ISS} = S_{\rm ISS}/V_{\perp}$ as the line of sight sweeps through scattering material at speed $V_{\perp}$.
A sharp pulse
at the pulsar will arrive over a range of times $\tau_{\rm ISS}$ at the observer;
the uncertainty principle relates $\tau_{\rm ISS}$ to the bandwidth of the scintillations $\Delta\nu_{\rm ISS} = 1/(2\pi \tau_{\rm ISS})$.
Over each element of the scintillation pattern $\Delta\nu_{\rm ISS}\times \Delta t_{\rm ISS}$, the propagation changes the electric field by a complex gain: a random amplitude and phase. 
At the source, the characteristic scale is $M S_{\rm ISS}$; a displacement of the source by this length has an effect 
that is statistically equivalent to displacement of the observer by $S_{\rm ISS}$.
The above basic parameters define the arena of our ``interstellar telescope'':
the high resolution of the scattering disk, acting as a lens, 
produces the statistical structure of the diffraction pattern in the observer plane,
which is then modified by the structure of the source.

The Vela pulsar is particularly attractive for statistical studies of interstellar scattering
because its scintillation bandwidth is relatively narrow at decimeter observing
wavelengths, so that many samples can be accumulated quickly.
The pulsar is strong, so the signal-to-noise ratio within one scintillation element is high.
The pulsar is scattered enough that the diameter of the scattering disk can be measured with Earth-based interferometry; 
this measurement allows us to determine the location of the scattering screen along the line of sight, and also allows,
in principle, studies of the two-dimensional structure of the source \citep{Gwi01,Shi10}.

\subsection{Pulsar Emission Structure via Interstellar Scattering}

A variety of authors have reported investigations of pulsar size using interstellar scattering.
The observables address either motion of the emission centroid, or the size of the emission region.
The present work falls into the second category.

Measurements of, or upper limits on, motion of the centroid of emission rely upon 
the reflex shift of the scintillation pattern in the observer plane,
when the pulsar rotates \citep{cor83,wol87,smi96,gup99}.
The observations measure such shifts from correlation of the scintillation spectrum at different phases of the pulse,
with spectra at later or earlier times.
Because motion of the source dominates changes in the scintillation pattern at the observer plane
over these timescales, 
such a correlation, with information on the location of the screen, yields the shift of the source.
These observations commonly find scales ranging from a few hundred to a few thousand km,
or up to the diameter of the light cylinder.

The phrase ``stars twinkle, planets do not'' expresses the fact that a finite source emission region can decrease the modulation by scintillation.
From the depth of modulation of scintillation,
one can infer the size of the emission region \citep{Coh66,Gwi98}.
This technique was used before the advent of synthesis interferometry to measure source sizes from scintillation in the interplanetary medium,
\citep{Rea72,Hew74}.
In ``strong'' scintillation, the modulation index $m=\sqrt{{{\langle I^2\rangle}/{\langle I\rangle^2}} - 1}$ of a point source is 100\% \citep{coh74}; for most pulsars at decimeter wavelengths, 
scintillation is very strong. Thus, modulation of less than 100\% would suggest the presence of source structure,
on the scale $M S_{\rm ISS}$.
\citet{Mac00} report an upper limit on the size of the Vela pulsar at $\lambda=45\ {\rm cm}$ observing wavelength, from the modulation index, although they do not discuss effects of intrinsic variability of the pulsar or self-noise;
these can also affect the modulation index.

We modify earlier approaches to measuring, or setting limits on, the size of the source from modulation, 
in that we fit a model to the distribution of intensity (or interferometric visibility).
Source size affects the smallest intensities or visibilities most strongly.
Intrinsic intensity variations and self-noise affect the largest intensities most strongly,
as we discuss in Section\ \ref{sec:instrumental_effects}.
All of these effects change the distribution function of intensity and visibility, and thus their moments.

The modulation index is a combination of first and second moments of intensity,
and thus is most sensitive to the largest intensities.
Thus, it is least sensitive to the shape of the distribution function where effects of source size are
largest, and most sensitive where they are smallest.
The full distributions of intensity or visibility 
provide more sensitive and complete information.
They can distinguish among effects that 
alter these distributions in different ways.

\subsection{Distributions of Electric Field, Intensity, and Visibility}

Scintillation affects the electric field of an astrophysical source at a particular frequency
by a gain and a phase \citep{Gwi11a},
here combined into the complex ``scintillation gain'' $\tilde g$.
For a point source in strong scattering, $\tilde g$ is drawn from a circular Gaussian distribution in the complex plane, 
with zero mean.
This behavior is a consequence of the large differences of path lengths, and the fact that many paths contribute to the 
signal received at the observer;
the Central Limit theorem then implies, under rather general assumptions, that the scintillation gain resulting from that sum over paths
is drawn from a Gaussian distribution.
This result is independent of assumptions about the distribution of scattering material; for example,
scattering in an extended or inhomogeneous medium can be treated quite generally via path integrals, and the same result holds \citep{flatte}.
In the time domain, the received signal is the emitted signal convolved with a kernel $g$ that
describes the scattering medium;
$g$ and $\tilde g$ form a Fourier transform pair \citep{Gwi11a}.

Because the scintillation gains are draws from a complex Gaussian distribution,
their square modulus is drawn from an exponential distribution.
An exponential distribution is thus the expected distribution of intensity for a scintillating point source, without effects of noise \citep{sch68,Gwi98}.
Similarly, the scintillation gains for two different antennas are drawn from correlated, complex Gaussian distributions.
Their product, the interferometric visibility, is drawn from the distribution of the product of such quantities.
This distribution is a zero-order modified Bessel function times an exponential \citep{Gwi01}.
In practice, to obtain these distributions the observer must average the intensity, or the interferometric visibility,
over many samples of the random electric field of the source.
This field is itself noiselike, and contributes to noise in the measurement via self-noise, as we discuss below.

If the source is extended, different points on the source will have different scintillation gains.
These gains decorrelate as the separation between points on the source plane increases.
A source consisting of two separated point sources provides a simple example.
If the two parts are separated by much less than $M S_{\rm ISS}$, then the gain factors are identical and the result is that for a point source above;
if the separation is much greater than $M S_{\rm ISS}$, then the observer records two superposed, independent scintillation patterns.
If the sources are incoherent, then the observed intensity is the sum of the two; and the distribution of observed intensity is the convolution of two exponential distributions.
The analogous results hold for interferometric visibility.
For a small, but extended source, the distributions of gains and phases for the different parts of the source are correlated;
however, they can be expressed as the convolution of the original distribution with distributions of the same form,
but of smaller scales \citep{Gwi01}.
As discussed in more detail below, finite size tends to concentrate the distribution of intensity near the mean,
and to soften the sharp cusp of the point-source distribution.

Noise  and intrinsic variations of flux density both broaden the observed distributions of intensity and interferometric visibility.
Noise includes contributions both from backgrounds and from the noiselike source.
Backgrounds are nearly independent of the flux density of the source, with small corrections
for quantization effects \citep{Gwi06,VelaNoise}.
Source noise has standard deviation proportional to the flux density of the source;
it is termed ``heteroscedastic,'' indicating that the variance is not constant \citep{Osl11}.
In combination, these contributions lead to variance of the noise given by a quadratic polynomial in phase with the signal,
and the linear terms of that polynomial at quadrature \citep{Gwi11b,VelaNoise}.
Intrinsic variations of flux density can be divided into 3 regimes according to time scale \citep{Gwi11b}.
The time to accumulate one sample of the spectrum is the 
product of the sampling rate and the number of spectral channels,
termed the ``accumulation time''.
Variations shorter than the accumulation time introduce correlations but do
not change the noise \citep{Gwi11a}.
Intermediate-term variations, longer than the accumulation time but shorter than the integration time,
contribute to noise.
Long-term variations, longer than the integration time, lead to a superposition of distributions with different
mean flux density.
Both noise and amplitude variations thus act in characteristic ways the can be distinguished from effects of size;
moreover, both broaden the distribution rather than narrowing it, as does finite size.

\subsection{Outline of Paper}

This paper focuses on our data and technique to estimate the size of the Vela pulsar's emission region from the distribution of
interferometric visibility. This measurement requires data with stationary instrumental gain, well-characterized noise, and rapidly-sampled, gated correlation.  Our analysis involves models for the effects of scintillation, noise, and amplitude variations of the pulsar.

In Section\ \ref{sec:observations}, we describe our observations of the Vela pulsar, correlation and gating using the DRAO VLBI correlator,
calibration, and fringing. 
We then present typical data and describe the formation of our data histograms.

Next, in Section\ \ref{sec:analysis}, we outline our analysis.
This analysis involves calculation of model histograms and fits to data.
We calculate the distribution of interferometric visibility in the complex plane,
for a small, circular source, in Section\ \ref{sec:vis_dist}.
We describe the distribution of noise in Section \ref{sec:noise}, and 
demonstrate how a superposition of distributions can model the pulsar amplitude variations in Section \ref{sec:amp_variations}.
We discuss the combination of these effects and evaluation of the model in Section\ \ref{sec:numerical}.
In particular, we demonstrate that the signature of finite source size is a W-shaped difference
of the best-fitting finite-size model from the best-fitting zero-size model.
We detail the numerical evaluation techniques in Section\ \ref{sec:model_calc},
and fitting techniques in Section\ \ref{sec:fitting}.

We present our results in Section\ \ref{sec:results_discussion}.
We first discuss an example fit in detail and show that the data histogram displays the characteristic W-shaped signature of source size.
We then give our results for all gates and spectral ranges.
We discuss various systematic effects that can contribute to, or bias, the inferred source size.
We briefly compare our results with previous results at $\lambda=13$\ cm,
and compare with other observational studies of the Vela pulsar's emission region.

In Section\ \ref{sec:summary}, we summarize our results.

\section{OBSERVATIONS, CORRELATION, AND CALIBRATION}\label{sec:observations}
\subsection{Observations}

We observed the Vela pulsar on
10 Dec 1997 using a network
comprising antennas at Tidbinbilla,
Mopra,
Hartebeesthoek,
and the VSOP spacecraft.
The observations began at 14:15\ UT and ended at 22:40\ UT,
for a time span of 8:25.
The observations were made at 18\ cm observing wavelength,
with left-circular polarization.
We recorded two frequency bands (IFs), of 16 MHz each, at each antenna.
The bands spanned 1634 to 1650\ MHz (IF1) and 1650 to 1666\ MHz (IF2).
Data were digitized (quantized and sampled) at recording time.

\subsection{Correlation}

The data were correlated with the Canadian S2 VLB correlator \citep{car99}.
This correlator is a reduced-table 4-level correlator.
Each IF was correlated separately with 8192 lags to form
a cross-correlation function.
The correlator was gated synchronously with the pulsar pulse,
in 6 gates across the pulse.
Each gate was 1\ msec wide.
The first 5 gates covered the pulse,
as shown in Figure\ \ref{fig:ampdraw}.
The sixth gate was located far from the pulse,
when the pulsar is ``off".
We averaged the results of the correlation for 2\ sec, or 22.4 pulsar periods;
except on the baselines to the spacecraft, which we averaged
for 0.5\ sec, or 5.6 pulsar periods.

\subsection{Editing and Fringing}

\subsubsection{Editing}\label{sec:editing}

The data were recorded in single sidebands;
thus, the spectra contained 8192 channels,
each with bandwidth 1.95\ kHz.
The cross-power spectra are complex.
The phase includes instrumental effects, primarily observational and instrumental
delays and rates \citep{tms86};
and effects of scintillation \citep{des92}.

We found that the time period from 19:10 to 21:13 UT on the Mopra-Tidbinbilla baseline contained data
with uniform high gain and low noise; absence of interference or gaps in correlation; and
little change in length and orientation of the baseline.
The coordinates projected perpendicular to the source direction vary over the range of $(u, v, w)$
from $(422.8, 1416.9, -392.67)$ to $(733.5, 1193.7, -614.5)\ \mu{\rm s}$. 
The projected baseline length is approximately 428\ km.

We edited the data to remove times and channels with interference or corrupt recording.
We identified channels that showed evidence of interference
such as high amplitude or high noise. This amounted to 27 channels for IF1 and 12 channels for IF2.
We also identified time records with excessively low or high amplitudes,
or with low correlation amplitudes, and removed those.

\subsubsection{Fringing}

We corrected for average delay and rate by fringe-fitting \citep[see][]{tms86}.
Delay represents a phase slope with frequency, and rate a phase slope with time.
We fringed the central 7168 channels of Gate 2, leaving ``guard zones'' 512 channels wide on each end, 
for the regions where passband gain rolled off and instrumental phase varied most rapidly.
We formed a 2-dimensional discrete Fourier transform to find the 
fringe rate for each 8-sample (16-sec) time interval,
using the traditional ``fringe'' algorithm \citep{tms86}. 

We then used the fringe rate and delay from Gate 2 to remove the corresponding phase slopes
from the other ``slaved'' pulsar gates.  
In tests with pairs of gates that contained strong signal, we verified that delay and rate
were the same for all gates, to the accuracy permitted by signal-to-noise ratio.
However, we found that the residual interferometer phase depended on pulsar gate.
We therefore used the phase estimated from each gate to correct that gate.
For this paper, the primary purpose of fringe-fitting was to remove 
instrumental phase, leaving only the effects of scintillation and those of
statistical noise in the data.

\subsubsection{Dynamic Cross-power Spectra}

The calibrated data take the form of complex cross-power spectra, sampled as a function of time, in 6 pulse gates and 2 IFs.
These data were gathered for several baselines, as discussed in \citet{VelaNoise}.
In this paper, we focus on the relatively short Mopra-Tidbinbilla baseline.
As an example, Figure\ \ref{fig:dynspec} shows the real part of the visibility,
for a short span of time and frequency in 3 gates.
The relative variation in amplitude between the gates has been removed by calibration, using 
the average amplitude over the spectral range for the time span of all the data,
as shown in Figure\ \ref{fig:ampdraw}.
The ratio of calibration factors was $24:36:13$ for gates 1, 2, and 3.  
The observed spectra differ because of the effects of noise, and because of variations in the relative amplitudes of individual pulses 
at the different pulse phases \citep{kri83,Joh01,Kra02,VelaNoise,Joh12a}.
The scattering medium is expected not to change between gates, which after all integrate over the same time interval;
thus, the scattering pattern should be the same, allowing for variations in noise and pulse-to-pulse variability.
An interesting question is whether differences might additionally reflect changes in the structure of the pulsar's emission region between gates.
Because the effects of noise and amplitude variations are random, 
this issue can only be treated statistically, using the correct descriptions of noise and amplitude variation.
Complicating this comparison is the fact that effects of source size are largest when scintillation leads to 
small flux densities \citep{Gwi98}.
The remainder of this paper makes such a statistical comparison.

\subsubsection{Histograms}\label{sec:histograms}

\begin{figure}[t]
\epsscale{.50}
\includegraphics*[width=0.98\textwidth]{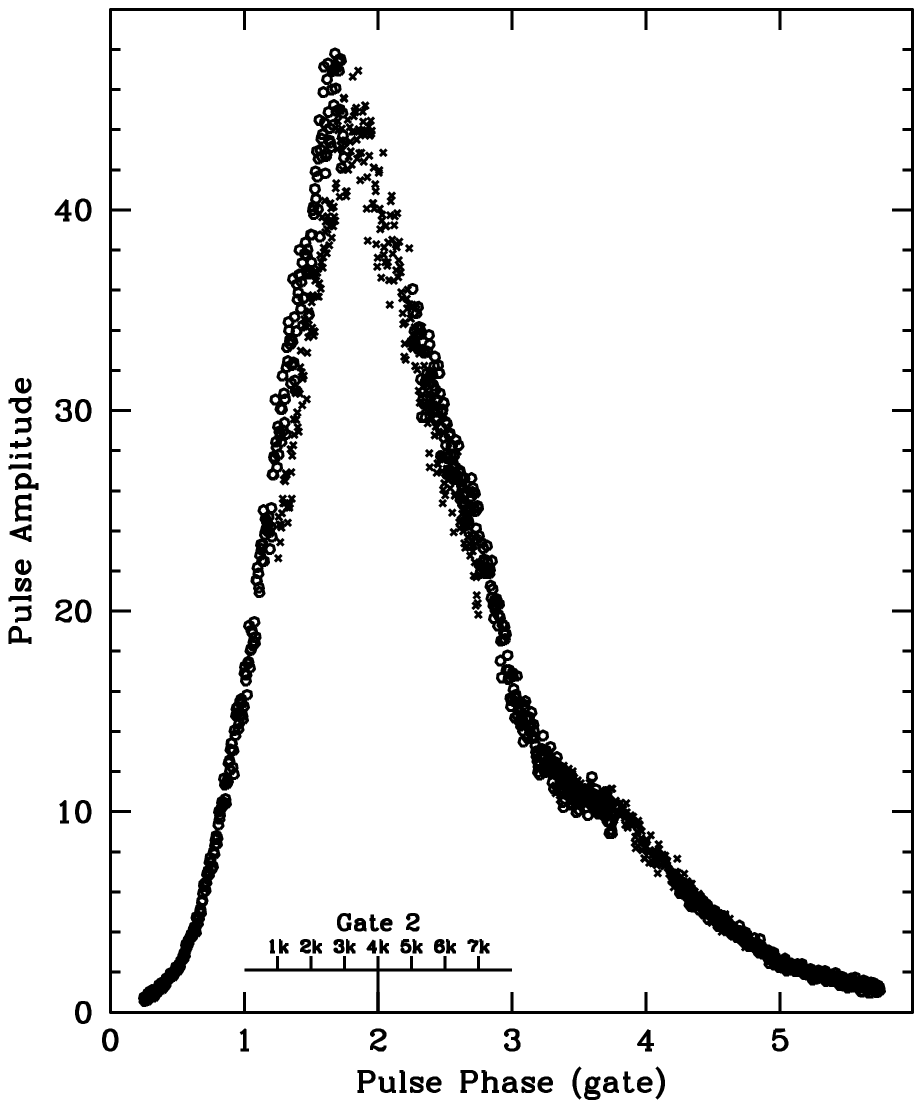}
\figcaption[]{
Pulse amplitude as a function of pulse gate and spectral channel.
Plotted is the average real part of the visibility for 
IF2, all gates and channels, from 19:10 to 21:13 UT,
boxcar-averaged over 20 spectral channels.
Gates and spectral ranges are indicated by alternating circles (odd) and crosses (even). 
Because of dispersion, spectral channel corresponds to pulse phase within a given gate; successive gates are offset by the gate width of 1\ ms.
Scale shows centers of channel ranges in multiples of 1024 for Gate 2.
The different gains, from changes in electric-field variance with fixed quantizer thresholds,
broaden the curve where gates overlap.
\label{fig:ampdraw}}  
\end{figure}

We reduce the observational data to a histogram of measurements of the real part ${\mathcal P}_N$,
and a histogram weighted by the mean square imaginary part ${\mathcal Q}_N$.
The subscript ``N'' denotes that these
distributions reflect measured histograms rather than model distributions.
Mathematically, these histograms correspond to sums over the observed points $V(\nu,t)$,
restricted to bins of width $w$ about the bin centers, at real part $X_k$:
\begin{align}
\label{eq:observed_projections}
{\mathcal P_N}(X_k) =& \sum_{\nu,t} 1                            &{\rm for\ } X_k - w/2 \leq {\rm Re}[V(\nu,t)] < X_k +w/2 \\
{\mathcal Q_N}(X_k) =& \sum_{\nu,t} {\rm Im}[V(\nu,t)]^2&{\rm for\ } X_k - w/2 \leq {\rm Re}[V(\nu,t)] < X_k +w/2  .
\nonumber
\end{align}
We sought to make the histograms from sufficiently narrow spectral ranges so that the amplitude did not vary greatly across the spectrum,
while including enough points for robust statistics.
Most of the spectral variation arises from the shape of the pulse and pulse dispersion,
as Figure\ \ref{fig:ampdraw}, and Figure\ 1 of \citet{VelaNoise}, suggest.
We adopted spectral ranges of 1024 channels within each 8192-channel spectrum from each gate.
We dropped the first and the last 1024 channels, to avoid effects of gain rolloff and phase variations near the edges of the observed band.
Consequently, our data are indexed by IF number (1 or 2), by gate (1 through 5), and by channel range in increments of 1024, beginning at 1024 to 6144.
Figure\ \ref{fig:ampdraw} shows the centroids of the 1024-channel ranges for Gate 2.

Figure\ \ref{fig:observed_distributions} shows examples of the measured histograms,
for IF1, gate 1, and channels 4096-5120.
We use these distributions, and others like them for other spectral and gate ranges,
to fit for the parameters of our theoretical models, as discussed in Section \ref{sec:analysis}.
We used intervals of $w = 0.0002$ 
for the histograms in this paper.  
Wider bins would average over the smooth, rapid variations of the distribution among bins.

Projection has a number of advantages.
One-dimensional distributions are much easier to visualize and fit than 2-dimensional distributions.
Projection increases the number of samples per cell, reducing Poisson noise. 
For the short Mopra-Tidbinbilla baseline, amplitude variations from scintillation affect primarily the real part, broadening it along the real axis.
Phase variation from scintillation affects the imaginary part,
and noise affects both real and imaginary parts.
Consequently, the second moment of the imaginary part in ${\mathcal Q}$ 
provides a useful constraint on noise and the effects of finite baseline length.
For a short baseline, visibility is concentrated near the real axis, so most of the information is contained
in these two distributions.

\section{ANALYSIS}\label{sec:analysis}
\subsection{Overall Strategy}

We seek to compare the observed distribution of interferometric visibility,
as expressed by the projected histograms, with theoretical models.
These models include the expected distribution of visibility for a scintillating source \citep{Gwi01}
and the distribution of noise \citep{Gwi11b,VelaNoise}.
Fitting a model involves calculating the model distribution for a given set of parameters,
quantifying its difference from the observed distribution with a figure of merit,
and then searching out the minimum such difference as a function of model parameters.
We wish to ensure that our search is as broad as possible,
so we begin with a grid search using the easily-calculated model for zero baseline length.
We then use the best-fitting parameters from this search as initial conditions for a fit using a baseline of finite length.
For both searches, we compare results for fits to a point-source model.
We discuss the signature of finite size in the histograms. 
We present the results of fits for the size of the source, as a function of pulse gate and frequency.
We discuss potential sources of  error, and the statistical significance of the result.
We compare these results with the inferred geometry of the pulsar's magnetic field.

\subsection{Distribution of Visibility}\label{sec:vis_dist}

\subsubsection{Background}

In the absence of noise or amplitude variations, the interferometric visibility for a scintillating point source is 
the product of complex Gaussian random variables. 
The resulting distribution of visibility peaks sharply at the origin;
indeed, the distribution is singular at that point \citep{Gwi01}.
Furthermore, the distribution has strong exponential wings that extend to large values.
This large dynamic range complicates evaluation.

For an extended source, also in the absence of noise or amplitude variations, 
the peak of the distribution softens and shifts toward greater real part, relative to that for a pointlike source.
However, the variance decreases in both the real and imaginary directions.
Thus, estimation of source size requires accurate calculation of the structure near the origin
on small scales, and of the behavior of the wings at large values. Figure\ \ref{fig:size_nosize_nonoise_distribution} shows distributions for sources with zero size and with small size, on a short baseline, calculated using the results of this section; 
the parameters are comparable to those we find for the Vela pulsar in Section\ \ref{sec:results_discussion} below.
We discuss effects of noise in Section\ \ref{sec:noise}, and effects of amplitude variability in Section\ \ref{sec:amp_variations}.

\subsubsection{Visibility of a Pointlike, Scintillating Source}

Explicitly, the normalized distribution of visibility for a pointlike, scintillating source,
without noise and with constant amplitude, is given by
\citep{Gwi01}:
\begin{equation}
P(V)=\frac{2}{\pi \kappa_0^2}\frac{1}{(1-\rho^2)}K_0\left(\frac{2}{(1-\rho^2)}\frac{|V|}{\kappa_0}\right)\text{exp}\left(\frac{2\rho}{(1-\rho^2)}\frac{\mathrm{Re}[V]}{\kappa_0}\right).
\label{eq:Pdist}
\end{equation}
Here, $\rho$ is the normalized covariance (that is, the normalized interferometric visibility of the source),
and $\kappa_0$ is the scale.
This distribution is that of a product, $x y^\ast$, of correlated, complex Gaussian random variables. It describes the distribution of visibility, after averaging over an ensemble of electric field measurements; it does not include effects of background noise, self-noise, or intrinsic source variability.

For a scintillating source viewed through an isotropic scattering screen,
\citep[][Equations 2, 4, 6]{Gwi01}
\begin{equation}
\rho = \exp\left\{ -{\frac{1}{2}} {\frac{(2 \pi)^2}{8 \ln 2}} {\frac{\theta_{\rm H}^2 b^2}{\lambda^2}} \right\} .
\label{eq:rhodef}
\end{equation}
Here, $\theta_{\rm H}$ is the full width at half maximum of the scattering disk, $b$ is the baseline length, 
and $\lambda$ is the observing wavelength.
The mean correlated flux density of a pointlike source is, therefore,
\begin{equation}
\langle V\rangle = \kappa_0 \rho = \kappa_0 \exp\left\{ -{\frac{1}{2}} (k \theta b)^2 \right\}.
\end{equation}
Here, the wavenumber is $k=2\pi/\lambda$, and the angular broadening is $\theta = \theta_{\rm H}/\sqrt{8\ln 2}$.
Angular brackets $\langle ... \rangle$ denote an average over many scintillations.

For short baselines, $\rho \approx 1$, and the distribution of visibility is concentrated near the positive real axis;
the imaginary part for a given value of ${\rm Re}[V]$ has variance that scales proportionately with ${\rm Re}[V]$.
For longer baselines, $\rho\rightarrow 0$, and the distribution is circularly symmetric about the origin.
Figure 2 of \citet{Gwi01} shows examples.

At zero baseline, $\rho\equiv 1$, visibility $V$ becomes intensity $I$, 
and $P(V)$ becomes the well-known exponential distribution of intensity for a scintillating point source:
\begin{equation}
P(I) = {\frac{1}{I_0}} \exp\{ - I/I_0\}.
\label{eq:expdist}
\end{equation}
To connect the two distributions, note that a ``zero-baseline interferometer'' can, in principle, measure complex visibility,
and will have complex noise; 
however, the visibility of the source is confined to the real axis, even with scintillation.
Thus, Eq.\ \ref{eq:expdist} is the projection of Eq.\ \ref{eq:Pdist} onto the positive real axis,
in the limit $\rho\rightarrow 1 $.

\subsubsection{Visibility of an Extended, Scintillating Source}

The distribution of visibility for an extended scintillating source,
without noise and with constant amplitude, is the convolution of a number of copies of Eq.\ \ref{eq:Pdist} \citep{Gwi01}.
For a small, scintillating source, the distribution is the convolution of 3 such copies,
two of them with scales $\kappa$ related to the size of the source along two orthogonal axes $\xi$, $\eta$:
\begin{eqnarray}\label{eq:kappa1s}
\kappa_{\xi} &=& \kappa_0 (k M \theta \sigma_{\xi})^2 \\
\kappa_{\eta} &=& \kappa_0 (k M \theta \sigma_{\eta})^2 \nonumber .
\end{eqnarray}
Here, $M = D/R$ is the effective magnification of the scattering disk,
where $D$ is the distance from observer to scatterer, and $R$ is the distance from scatterer to source.
The dimensions of the source are $\sigma_{\xi}$, $\sigma_{\eta}$.
We assume that the source is small in the sense that $(k M \theta \sigma_{\xi}), (k M \theta \sigma_{\eta}) << 1$.
If this condition does not hold, then additional terms are important; the convolution involves more distributions.
The covariances, corresponding to $\rho$, change for the subsidiary distributions as well:
\begin{eqnarray}\label{eq:subsidiary_rhos}
\rho_{\xi} &=& \left(1-(b_{\xi} k \theta)^2 \right) \exp\left\{ -{\frac{1}{2}} (k \theta |b|)^2 \right\} \\
\rho_{\eta} &=&  \left(1-(b_{\eta} k \theta)^2\right) \exp\left\{ -{\frac{1}{2}} (k \theta |b|)^2 \right\}  \nonumber .
\end{eqnarray}
These are nearly equal to $\rho$ for short baselines.

We have not found a simple analytic expression for the result of the convolution for $\rho<1$.
Moreover, the convolution is challenging to reproduce numerically,
because $P(V)$ has both a sharp peak at the origin, and high skirts that extend to large $|V|$.
Our strategy is therefore to proceed as far as possible via analytic calculations,
and then evaluate the remaining integrals numerically.

%
To achieve our first numerical reduction, note that we can easily convolve visibilities that are drawn from identical distributions. 
The average of $N$ such visibilities is distributed according to
\begin{eqnarray}
P_N(V) = \frac{2}{\pi} \frac{N^{N+1}}{(N-1)!} \frac{1}{\left(1-\rho^2 \right)\kappa^2} \left(\frac{|V|}{\kappa} \right)^{N-1} K_{N-1} \left(\frac{2 N}{1-\rho^2} \frac{|V|}{\kappa} \right) \exp\left(\frac{2 N \rho}{1-\rho^2} \frac{\mathrm{Re}[V]}{\kappa} \right).
\end{eqnarray}
We can use this result to calculate the convolution of visibilities drawn from different distributions if we first use Feynman parameters \citep{sred} in Fourier space to symmetrize the corresponding conjugate product of functions. Because the visibility is complex, convolution of $N$ visibilities requires a $2(N-1)$-dimensional integral. However, to symmetrize the convolution requires a single Feynman parameter for each visibility. The Feynman parameters also have an overall $\delta$-function constraint, so the convolution is reduced to an $(N-1)$-dimensional integral. 

For a small, circular source, the observed distribution is a convolution of three visibilities, two of which are drawn from identical distributions.  These two distributions are parametrized by 
Equations\ \ref{eq:kappa1s} and\ \ref{eq:subsidiary_rhos}
with $\sigma_{\xi}=\sigma_{\eta}$ and $b_{\xi} k \theta<<1$, $b_{\eta} k \theta <<1$.
This symmetry allows elimination of an additional degree of freedom, so the original four-dimensional visibility integral is reduced to a one-dimensional integral. Explicitly, we assume that $\kappa_0$, $\rho_0$, $\kappa_1$, and $\rho_1$ are arbitrary, but $\kappa_2=\kappa_1$ and $\rho_2=\rho_1$. For convenience, we also introduce parameters $a_i \equiv \kappa_i\left(1 - \rho_i \right)/2$ and $p_i \equiv 2\left[\pi \kappa^2 \left(1-\rho^2\right) \right]^{-1}$. Then, the distribution of visibility is given by
\begin{eqnarray}
\label{eq:Pconv}
P\left(V\right) &\equiv & \left(P_0\ast P_1 \ast P_2\right)\left(V\right) \\
\nonumber &=& \pi^{2} p_0 p_1^2 a_0^2 a_1^4 |V|^2 \int_0^1 ds \ (1-s)f_1(s)K_{2}\left(f_2(s)\left|V\right|\right)e^{f_3(s) Re[V]},
\end{eqnarray}
where we have defined, for convenience, the 3 functions
\begin{eqnarray}
f_1(s) &\equiv & \left[\left(a_0^2 s + a_1^2(1-s)\right)\left(a_0^2 s + (1-s)\left(a_1^2 - s \left[a_1\rho_0-a_0\rho_1\right]^2\right)\right)\right]^{-1}\\
\nonumber f_2(s) &\equiv & \frac{\sqrt{a_0^2 s + (1-s)\left(a_1^2 - s \left[a_1\rho_0-a_0\rho_1\right]^2\right)}}{a_0^2 s + a_1^2(1-s)}\\
\nonumber f_3(s) &\equiv & \frac{a_0 \rho_0 s + a_1 \rho_1 (1-s)}{a_0^2s+a_1^2(1-s)}.
\end{eqnarray}
This one-dimensional integral is suitable for efficient numerical evaluation.

\subsection{Noise}\label{sec:noise}

Noise broadens the distribution of visibility described in the previous section. It softens the peak without shifting the centroid of the distribution.
Noise arises as a background from unrelated sources and within the telescopes, and from the source itself (``self-noise'').
In principle, background noise is independent of the behavior of the source,
whereas self-noise increases with the flux density of the source,
and has different magnitude in and out of phase with the interferometric signal \citep{VelaNoise}.

The Dicke equation conveniently describes noise and self-noise.
This equation states that 
error in measurements of antenna temperature $\delta T$ varies with total system temperature $T$,
including the contribution of the source \citep{Dic46}:
\begin{equation}
\label{eq:dicke_eq}
(\delta T)^2 = {\frac{T^2}{ N_{\rm obs} }} ,
\end{equation}
where $N_{\rm obs}$ is the number of samples.
This equation
describes how accurately $N_{\rm obs}$ observed samples from a Gaussian distribution 
can measure the variance of the distribution (or, for interferometry, the covariance of two distributions).
The net noise in interferometric visibility has variance that increases quadratically with the signal in phase with the signal;
and linearly, with the same constant and linear coefficients, at quadrature with the signal \citep{Gwi11b,VelaNoise}.

The effect of noise on the distribution is not a convolution, because the noise depends on the visibility. Hence, 
it resembles a convolution, but with non-stationary kernel.
Because we average over approximately 33 elements in frequency and time for each sample, after taking bandwidth, integration time, spectral resolution, and pulse gating into account, we assume that the noise follows a Gaussian distribution.
For interferometric observations, a quadratic polynomial specifies the variance of noise in phase with the visibility, 
$\sigma_{||}^2$;
and the linear terms specify $\sigma_{\perp}^2$, the variance at quadrature:
\begin{eqnarray}
\label{eq:noise_polynomial_complex} 
\sigma_{||}^2 = \langle (\delta {\rm Re}[V])^2\rangle &=& b_0 +b_1\langle |V|\rangle + b_2 \langle |V|\rangle ^2 \\
\sigma_{\perp}^2 = \langle (\delta {\rm Im}[V])^2\rangle &=& b_0 +b_1\langle |V|\rangle.
\nonumber
\end{eqnarray}
The parameters $\{ b_0, b_1, b_2\}$ describe the noise.
For a source of constant flux density, $b_2$ is simply the reciprocal of the number of samples, $N_{\rm obs}$.
For an source of varying amplitude,
$b_2$ is $1/N_{\rm obs}$ plus $(\delta I/I)^2$. Here $\delta I$ is the 
standard deviation of the flux density on timescales longer than the accumulation time $8192/16{\rm\ MHz} = 512\ \mu{\rm sec}$,
but shorter than our 2-sec integration time
\citep[see][Section 2.2.2]{Gwi11b}.

Quantization of the signal, when it is digitized for recording, also affects noise.
Quantization introduces additional background noise, and a gain that scales signal and noise \citep{Gwi06}.
These parameters change with quantizer levels, in units of the variance of the electric field.
Consequently, for this experiment they 
change with pulse gate \citep{VelaNoise}.

In this work, we fitted for the coefficients $\{b_0,b_1,b_2\}$ in Equation\ \ref{eq:noise_polynomial_complex}, for each pulse gate and spectral range.
We fit $b_0$ and $b_1$ freely, and fit for $b_2$ 
with the requirement that $b_2\geq 1/N_{\rm obs} = 1/33$.
The results of these fits are comparable with the noise parameters obtained from differences of samples adjacent in time
in \citet{VelaNoise}, when the signal-to-noise ratio provides differences with enough accuracy
to measure the noise.
The binning technique of \citet{VelaNoise} provides for easy visualization of the noise distribution
and is independent of the underlying distribution of visibility,
but is subject to biases and requires high signal-to-noise ratio.
We adopt the global-fitting technique described above for this work, because it is more precise and is applicable for arbitrary
signal-to-noise ratio.

\subsection{Amplitude Variations}\label{sec:amp_variations}

The pulsar changes amplitude with both time and frequency during the observations.
For our observations, variations in shape and amplitude of individual pulses dominate time variability.
Individual pulses vary in flux density,
as well as in shape and arrival time \citep{kri83,Joh01,Kra02}.
For the Vela pulsar, these changes are almost uncorrelated between successive pulses. 
Variability changes with pulse phase: 
it is largest at the beginning and end of the pulse, and less during the pulse.

Because the pulse is dispersed,
higher-frequency channels sample a later pulse phase than lower-frequency.
Thus, the average spectrum reflects the pulse profile, as shown in Figure\ \ref{fig:ampdraw}.
Dispersion produces significant amplitude variations with frequency, even over a range of 1024 channels.
These spectral variations reflect the average shape of the pulse, so they are stationary with time.
Comparison of amplitudes averaged over a 1024-channel range of data,
over longer timescales, show no further, slow effect of gain variations or variations of the source over
the span of our data, as described in Section\ \ref{sec:observations}.

Intrinsic amplitude variations on timescales shorter than the accumulation time do not affect the distribution of noise \citep{Gwi11b}.
On timescales between the accumulation time and the integration time,
amplitude variations contribute to noise through $b_2$ \citep{Gwi11b}.
On timescales longer than the 2-sec integration time,
intrinsic amplitude variations lead to superposition of distributions, 
with different values for the amplitude parameter $\kappa_0$ \citep{Gwi00}.
Our 2-sec integration averages over 22 or 23 pulses,
reducing the expected modulation by a factor of approximately 4.7.
We parameterize the remaining amplitude variation
by the intrinsic modulation index $m_{\rm s}=\sqrt{\langle I^2\rangle_{\rm s2}/\langle I\rangle_{\rm s2}^2 - 1}$,
where the subscripted angular brackets $\langle ... \rangle_{\rm s2}$ indicate an average over those intrinsic variations
with timescales longer than the integration time of 2\ sec.

\subsection{Evaluation}\label{sec:numerical}

\subsubsection{Projection onto the Real Axis}

We project the model distributions, including effects of noise and amplitude variations, onto the real axis.
We calculate the projected probability density, 
and the summed squared imaginary part:
\begin{eqnarray}
\label{eq:projections}
\mathcal{P}(X_k) &=& \int_{X_k-w/2}^{X_k+w/2} d{\rm Re}[V]\;  \int_{-\infty}^\infty d{\rm Im}[V]\; P (V) \\
\mathcal{Q}(X_k) &=&  \int_{X_k-w/2}^{X_k+w/2} d{\rm Re}[V]\;  \int_{-\infty}^\infty d {\rm Im}[V]\; {\rm Im}[V]^2\;  P (V) .
\nonumber
\end{eqnarray}
These are scaled models for the projections ${\mathcal P_N}$, ${\mathcal Q_N}$ of the data as described in 
Section\ \ref{sec:histograms}, and plotted in Figure\ \ref{fig:observed_distributions}.
Note that $\mathcal{Q}(X_k)$, the mean square imaginary part in each bin, is a second moment of ${\rm Im}[V]$, and so is expected
to be noisier than $\mathcal{P}(X_k)$, the zeroth moment of ${\rm Im}[V]$.
The higher noise for ${\mathcal Q}$ in Figure\ \ref{fig:observed_distributions} reflects this.

\subsubsection{Noise and Amplitude Variations}\label{sec:noiseandamplitudevariations}

Effects of noise and amplitude variations must be calculated
numerically for the full two-dimensional distribution,
and then projected. 
Noise changes the value of each measurement,
and thus spreads the corresponding probability distribution function over a surrounding region.
It smoothes a spike, corresponding to a single deterministic value,
into an elliptical Gaussian distribution centered at that point,
with variances given by the noise polynomial, Equation\ \ref{eq:noise_polynomial_complex}, evaluated at that point.
The effect of noise on the distribution of visibility 
thus resembles a convolution of the probability distribution of the scintillating source,
with the Gaussian distribution of noise. It is not a convolution,
because the noise depends on the signal:
this operation is sometimes called a ``convolution with varying kernel''.
The Gaussian noise, with dependence on signal strength, is the kernel in this case.
Because the kernel varies, we must project after convolution.
Because the distribution is relatively concentrated along the real axis, the notion of convolution after projection has intuitive value,
but we do not use this construction in our analysis for $\rho<1$;
we calculate the full variation of the noise kernel.

The effect of amplitude variations on timescales longer than the integration time is to superpose distributions with different $\kappa_0$ but the same noise polynomial.
We implement this effect numerically, by averaging a number of distributions with varying $\kappa_0$. 
Fortunately, integration over 2 sec reduces the intrinsic amplitude variations, as discussed below.

\subsubsection{Sample Evaluation}\label{sec:modelplot}

A sample calculation illustrates our method, and shows the origin of the W-shaped signature of finite source size
in the difference of projected distributions $\mathcal{P}$.
Figure\ \ref{fig:projected_model_distributions} shows how the best-fitting zero-size and finite-size models differ, for the span of data shown in 
Figure\ \ref{fig:size_nosize_nonoise_distribution}.
The upper panel of the figure shows the projected distribution
$\mathcal{P}$ without noise or amplitude variations.
For zero size, the projected distribution shows a sharp cusp at ${\rm Re}[V]=0$,
with an exponential decline along the positive real axis of $V$,
and a more sharply-declining exponential along the negative real axis.
The negative-side exponential declines so sharply because $\rho=0.986\approx 1$ in this example.
For a model with finite size, with the same normalization and mean,
the bulk of the projected distribution is shifted toward the positive real axis,
with a rounded peak.  
For the same mean amplitude, the peak has a narrower spread.

Note that the point-source model has greater probability at large and small ${\rm Re}[V]$
than does the extended-source model,
but the extended model has greater probability in between.
This behavior arises because we require that the two distributions have the same mean and normalization.
The shift of the maximum toward greater ${\rm Re}[V]$ for the finite-size model
then requires a reduction of $\kappa_0$
\citep[see][Equation 31]{Gwi01}.
The largest exponential scale, $\kappa_0$, dominates the behavior of the distribution away from the origin,
so that a finite-size source concentrates the distribution of visibility.

Noise broadens the distributions, as the middle panel in Figure\ \ref{fig:projected_model_distributions} illustrates.
The best-fitting noise parameters are $\{ b_0, b_1, b_2\} = \{ 0.000062, 0.0037, 0.078\}$ for the zero-size model, and $\{ 0.000058, 0.0045, 0.181\}$ for finite-size. 
The noise at $V=0$ has standard deviation $\sqrt{b_0}$, or approximately 0.008.
This value is comparable to the width of the distribution,
so that the details apparent in the top panel of the figure are blurred.
Moreover, the noise increases away from $V=0$, as given by the higher-order coefficients.

Intrinsic amplitude modulation also changes the distributions.
The mean square of the intrinsic amplitude modulation is 
$m_{\rm s}^2= 0.009$ for the best-fitting zero-size model, and
$m_{\rm s}^2 = 0.007$ for the best-fitting finite-size model,
for our example of spectral and pulse-gate range.
In both cases the degree of modulation is small as expected after integrating 22 or 23 pulses.

The lower panel in Figure\ \ref{fig:projected_model_distributions} shows the difference of the zero-size and finite-size models.
Even after including effects of noise and variability, the underlying features of the two distributions persist:
the finite-size distribution has smaller probability density for ${\rm Re}[V]<0$, 
because of its lesser density near ${\rm Re}[V]=0$ in the top panel;
and at large ${\rm Re}[V]$, because of its more rapid decline there.
It has greater density near the mean amplitude.
Thus, we expect the signature of a finite-size source to be a W-shaped difference of ${\mathcal P}$ for 
the best finite-size model from the best 
zero-size model,
after effects of noise and intrinsic amplitude variations have been included.

The situation for the distribution of mean square imaginary part $\mathcal Q$ is somewhat similar to that for $\mathcal P$,
but the fractional difference between models is less.
Indeed, nearly all of the spread in imaginary part results from noise, 
for values of $\rho$ of interest for this baseline;
so that the distribution $\mathcal Q$ functions more as a constraint on the noise model,
than as a carrier of information about source size.
We present plots for $\mathcal Q$ below, in Section\ \ref{sec:pox}.

\subsection{Calculation of model}\label{sec:model_calc}

\subsubsection{Nested Iterative Integration}

Our calculation of the model distribution involves 4 nested iterative loops.
We calculate $\mathcal{P}$ and $\mathcal{Q}$ in parallel, on a grid of points in the real part of visibility, $X_k$.
Each histogram bin has width $w = 2\times 10^{-4}$.
This size is narrow enough to track the behavior of the distribution accurately,
but wide enough to contain enough visibilities so that the Poisson noise does not obscure the model differences.

At the lowest level, we integrate over $s$ to find the probability density $P(V)$ at a point $V$
in the complex plane, using Equation\ \ref{eq:Pconv}.
We integrate this expression iteratively using Simpson's rule \citep{NumRec} with $2\times 3^N$ segments of equal width on the $N^{\text{th}}$ iteration.
The integration terminates when successive iterations have a fractional difference of less than $0.1\%$, or $N>10$. Because the error bound for this integration scheme goes as $N^{-4}$, this criterion yields an expected accuracy of $\sim0.001\%$ -- a small fraction of the magnitude of measurable source size effects.

We broaden the probability density at each grid point by its 
corresponding elliptical Gaussian distribution of noise.
The polynomial coefficients $\{ b_0, b_1, b_2\}$  (Equations\ \ref{eq:noise_polynomial_complex}) parametrize the noise, and the complex visibility at the grid point determines the scale and orientation of the ellipse.
We use the analytic expressions for the projection of a Gaussian distribution and the projected second moment of a Gaussian
distribution to evaluate the contribution of each point, with noise,
to the projections $\mathcal{P}$ and $\mathcal{Q}$ on our grid of points.

At the second level, 
we integrate the contributions to $\mathcal{P}$ and $\mathcal{Q}$ over the imaginary part of $V$.
We integrate from the real axis out to the 4-sigma standard deviation for the distribution
in the absence of size effects, as calculated from the second moment of Equation\ \ref{eq:Pdist}.
Again, we use Simpson's rule iteratively with $2\times 3^N$ segments of equal width, but now require a fractional change of less than $10^{-8}$ in $\mathcal{P}(X_k)$ between two successive iterations.
We monitor $\mathcal{P}(X_k)$ because it is more sensitive than $\mathcal{Q}(X_k)$ to behavior near the real axis, where the 
distribution varies most rapidly, 
so that $\mathcal{P}(X_k)$ converges more slowly.

At the third level, 
we integrate the values of $\mathcal{P}$ and $\mathcal{Q}$ over sub-bins within each histogram bin along the real axis. 
This step accounts for effects of finite histogram bin width in discrete representations of a probability distribution function. 
It is most important when there is large-amplitude non-linear structure, as exhibited by the rapid rise of $P$ at the origin and the cusp of the noise-free distribution near the origin. In the limit of small bin width, $w$, the histogram representation of a function $f(x)$ will systematically overestimate the function at $x_0$ by $w^2 f''(x_0)/24$. We set the number of integrated sub-bins to 3;
comparison with computations using 5 and 7 sub-bins for the fits that obtained smaller sizes yielded insignificant differences.

At the fourth and highest level, we account for the intrinsic amplitude variability of our 2-sec integrations.
We calculate each visibility distribution 5 times, with different source amplitudes, as given by a Gaussian distribution 
centered at 1, scaled by an amplitude-variation parameter.  We then superpose these 5 distributions to
produce a distribution including effects of variations of amplitude of the source on the final distribution.
The procedure is that used by \citet{Gwi00} to describe the distribution of scintillation in the presence of amplitude variations.
We found no difference when using more finely-divided distributions of source amplitude.
Effects of amplitude variations would be significant if the amplitude approached zero;
but
because Vela does not null and variations of individual pulses are only weakly correlated,
amplitude variations after 2-sec integrations are small.

\subsubsection{Exponential model}\label{sec:exponential_fits}

For zero baseline, $\rho=1$, the distribution of visibility is a sum of exponentials on the positive real axis
and zero elsewhere \citep{Gwi98}.
This form obviates the lowest 2 levels of the 4-fold nested loops above,
so the model is faster to calculate and, thus, is useful for exploration of parameter space.
For this exponential model,
we assumed a circular source,
as we do for our $\rho<1$ calculation.
Tests showed that the two calculations were equivalent for $\rho\rightarrow 1$: a useful check of 
our numerical integration.
		
\subsection{Fitting}\label{sec:fitting}
	
\subsubsection{Fit Parameters}\label{sec:fit_parameters}

Seven parameters describe the model.
Three of these are the coefficients of the noise polynomial, Equation\ \ref{eq:noise_polynomial_complex}: $\{ b_0, b_1, b_2\}$. 
A fourth and fifth describe the distribution in the absence of noise and amplitude variations:
the amplitude scale $\kappa_0$ and the normalized mean interferometric visibility $\rho$.
The scale $\kappa_0$ is proportional to flux density.
The mean interferometric visibility $\rho$ describes the baseline length and angular size of the scattering disk 
(Equation\ \ref{eq:rhodef}).
We do not provide a parameter for overall normalization: 
we constrain the normalization of the model distribution to equal that of the observed distribution.
A sixth parameter describes intrinsic amplitude variations $(m_{\rm s}^2)=(\delta I^2/{\langle I\rangle}^2)$
on timescales longer than an integration time;
because of such variations, the observed distribution involves a superposition of different values
of $\kappa_0$.
The parameter $(m_{\rm s}^2)$ gives the variance of this distribution of amplitude fluctuations, normalized by the mean amplitude.
A seventh parameter gives the size of the source,
usually expressed as $\kappa_{\rm ratio} = \kappa_1/\kappa_0 = (k M\theta\sigma)^2$.
In principle, additional parameters give correlations $\mu$, $\nu$ for the subsidiary distributions due to source structure  and the intrinsic elongation of the source \citep[][Equation 28]{Gwi01};
however, these are expected to be nearly equal to $\rho$ for a short baseline.

Among the effects we did not include in our model are elongation of the scattering disk and elongation of the source.
For a constant orientation of the baseline, 
elongation of the scattering disk has no effect on the distribution of visibility, 
for the appropriate parameters $\rho$, $\kappa_0$, and $\kappa_{\rm ratio}$ \citep{Gwi01}.
The rotation of our baseline was small during the test interval considered here.
Effects of elongation of the source are more subtle, but appear in our simple model only when the source is
nearly resolved by the scattering disk, or the baseline is relatively long.  

Note that $\rho$ is a property of scattering, and should be constant over the entire pulse.
Finite source size will change it slightly \citep[][Equation 18]{Gwi01}.
A source with spatial coherence, over scales comparable to those detectable
via scintillation, can also change $\rho$ by illuminating only part of the scattering disk \citep{Gwi98}.

\subsubsection{Summed, Weighted Squared Residuals}\label{sec:combine2projections}

Our fits minimize the weighted mean square difference between the data and the model.
Individual bins in the histogram ${\mathcal P}$ 
measure a number of counts $N$, and are expected to display Poisson statistics.
For ${\mathcal Q}$ the values are squares of drawn from a nearly Gaussian distribution,
and should display statistics analogous to that of the Dicke equation, Equation\ \ref{eq:dicke_eq}.
In both cases, the distribution of noise is nearly Gaussian, with variance proportional to the bin value.
We adopt this weighting above a threshold of $N=100$ counts, with constant weighting below that value
so as to reduce the influence of bins with zero or few counts.

The two histograms ${\mathcal P}$ and ${\mathcal Q}$
have different dimensions and different vertical scales.
We seek to combine the residuals for both into one figure of merit.
A reasonable conversion factor is simply the quotient of the integrated areas under the distributions, which has the correct dimensions.
However, we expect the second moment to be noisier, as Figure\ \ref{fig:observed_distributions} shows.
To quantify this noise in ${\mathcal Q}$,
we fit the distributions with simple models to find smoothed average values, and then difference adjacent points
to determine the noise.
We find that the standard deviation of noise in ${\mathcal Q}$ determined in this way
is typically 3 times that in ${\mathcal P}$.  
We therefore weight the residuals in ${\mathcal Q}$ by $1/9$ of the ratio of the areas under the two histograms.
This has the effect of making the mean square residual roughly equal for the two, for our models. 

We find that the mean square residual is close to 1 for ${\mathcal P}$ with this weighting, 
for the fits described in Section\ \ref{sec:results_discussion} below.
This indicates that the statistics are indeed Poisson:
number of counts limits the accuracy of the histogram values in our narrow bins.
As expected for our scaled weighting,
the mean square residual is close to 1 for ${\mathcal Q}$, as well.

As a further test, we also fit to ${\mathcal P}$ and ${\mathcal Q}$ independently.
We find that fits to only ${\mathcal P}$ found minima close to those found using both 
${\mathcal P}$ and ${\mathcal Q}$, with equal or sometimes larger intrinsic size of the source. 
Fits to only ${\mathcal Q}$ typically do not converge; apparently this distribution does not contain enough information to determine
model parameters independently.
Fits with uniform rather than Poisson weighting yield similar results for size, but tend to converge less quickly.

\subsubsection{Grid Search Using Exponential Model}\label{sec:gridsearch}

In order to examine parameter space over large scales,
and to provide initial parameters for our fits for arbitrary $\rho$,
we performed a grid search using an exponential model, as described in Section \ref{sec:exponential_fits}.
Because such a model can be calculated quickly, a grid search can survey large regions of parameter space efficiently.
In our grid searches, 
we fit for three parameters: two noise coefficients $b_0$, $b_1$ and the amplitude $\kappa_0$.
We searched a grid of parameters in the noise coefficient $b_2$ and the size parameter $(k M \theta\sigma)^2$.
We demanded that $b_2 > 0.030$, as dictated by the number of samples correlated \citep{VelaNoise},
and searched $0.030 < b_2 < 0.4$.
We ignored effects of longer-term intrinsic amplitude fluctuations between integrations: $m_{\rm s}\equiv 0$.

The grid search indicated that the summed, weighted squared residuals vary smoothly over the parameter space defined by the five varying parameters.
We found only one minimum in all IF, spectral and gate ranges.
The best-fitting size of the source $\sigma$ 
defined by this minimum was usually significantly different from 0, with a reduction in the weighted residuals
from the best-fitting zero-size model
comparable to that found from the more sophisticated fits discussed below. The size parameter
$(k M \theta\sigma)^2$ usually agreed to within 30\% for the two
kinds of fits.
The amplitude-modulation parameter $m_{\rm s}$ is responsible for much of the difference in size.
The minimum found by the grid search provided a useful starting point for the much-slower fits with $\rho<1$ and $m_{\rm s}>0$,
discussed in Section\ \ref{sec:results_discussion} below.  

\subsubsection{Levenberg-Marquardt Algorithm}\label{sec:LM_fitting}

We 
use calculations of ${\mathcal P}$ and ${\mathcal Q}$ as described in Section\ \ref{sec:numerical} to
fit for model parameters with finite size and $\rho<1$, using the Levenberg-Marquardt algorithm \citep{NumRec}.
We fit for the 6 parameters described in Section\ \ref{sec:fit_parameters},
using weighting described in\ \ref{sec:combine2projections}.
We initialize these parameters using the results of the grid search with an exponential model. 

\subsubsection{Normalized Visibility Parameter $\rho$}\label{sec:rho}

We expect $\rho=0.986$,
based on the parameters reported by 
\citet{Gwi97}: angular broadening of $(3.3 \times 2.2)\  {\rm mas}$
(full width at half-maximum intensity), with the major axis at position angle $92^{\circ}$,
in observations at wavelength $\lambda=13\ {\rm cm}$.
We scale these to our observing frequency by $\theta\propto \nu^2$,
and use the length and orientation of our baseline as described in Section\ \ref{sec:editing}
to find $\rho$, using Equation 28 of \citet{Gwi01}.
In tests,
we found that the mean square residual for this data set was not particularly sensitive to $\rho$, for $0.85 < \rho < 1$,
so that our observations do not provide a good way to fit for $\rho$:
the baseline is too short to provide much information.
Our results for the fitted size, in particular, were not sensitive to $\rho$ within this range: using $\rho=0.92$ rather than $\rho=0.986$
changed the inferred size parameter $(k M\theta\sigma)$ by less than 10\%.
In several frequency ranges where the pulse was strong 
(channels 1024 through 4096 of Gate 2 of both IFs), fits including $\rho$ strongly favored $\rho>0.9$, with $\rho=0.92$ sometimes favored within
that range.  
We expect $\rho$ to be almost the same for all of the data, in all channels and gates.
Differences from changes in frequency were too small to detect.
Effects of source size are expected to be small for our short baseline.
Indeed, we expect from the moments of the distribution that $\rho$ and size parameter $(k M \theta\sigma)^2$ should have little
covariance \citep{Joh12c}.
We adopt $\rho=0.986$, noting that revisions may cause small changes in the inferred size.

\section{RESULTS AND DISCUSSION}\label{sec:results_discussion}

\subsection{Finite- and Zero-Size Fits}\label{sec:size_nosize}

We performed independent Levenberg-Marquardt fits for both finite and zero size sources.
Comparison of the two provides a measure of the significance of our results.
Both sets of fits were initialized with the best-fitting parameters for finite or zero size 
found from our grid search.
Both allowed all the other parameters to vary:
noise coefficients $\{ b_0, b_1, b_2\}$,
amplitude scale $\kappa_0$, and intrinsic amplitude variations on timescales longer than our integration time, $m_{\rm s}$.

\subsubsection{Sample Fit}

The lower panels of Figure\ \ref{fig:observed_distributions} show the residuals to a fit with zero source size,
in histogram form. Structure remains in the residual histogram, apparent as variations near zero visibility.
An independent fit, to the same data but with a finite source size, removes most of these variations.
To illustrate this, we display the difference of the finite-size model from the zero-size model,
as smooth curves. 
The difference curve for ${\mathcal P}$ is the same as that shown in the lower panel of Figure\ \ref{fig:projected_model_distributions}.
Like the residuals, this curve shows the W-shaped signature of finite source size.

The mean squared residuals in ${\mathcal P}$, after fitting for finite source size,
are approximately those expected for Poisson-noise dominated errors.
The data contain 437 bins with more than 100 counts.
The sum of the squared Poisson-weighted residuals, after fitting, is 434.
Thus, the mean square errors are approximately as expected.
The fit to a model for a source with zero size leads to a sum of the squared Poisson-weighted residuals of 621,
significantly greater.

For ${\mathcal Q}$, the sum of the squared residuals, after correction by the relative areas of ${\mathcal P}$ and ${\mathcal Q}$,
and the factor of 3 from comparison of differences, is 705 for a finite-size model, and 785 for a zero-size model.
The residuals errors in ${\mathcal Q}$ appear noiselike in the plot, suggesting that the factor of 3 estimated from differences 
in other spectral and gate samples may be 
low for this sample of data.
Nevertheless, the finite-size model improves the fit for ${\mathcal Q}$.

The reduction in summed mean-square weighted residuals for ${\mathcal P}$ and ${\mathcal Q}$ together is 19\%.
This difference is highly statistically significant at more than the
40-$\sigma$ level, for our sample size and number of parameters, according to the F-ratio test \citep{Bev}.
At this high level,  finite sampling limitations are
unlikely to dominate errors in our estimates of emission size. 
However, this figure demonstrates that the signature
of finite size appears in the data, as inspection of the figures suggests.
The best-fitting size parameter is $(k M \theta \sigma)^2 = 0.0423$,
or scaled size $(k M \theta \sigma) = 0.21$.
This corresponds to a source size of approximately 180\ km (standard deviation of the Gaussian distribution), 
or to approximately 420\ km for the full width at half-maximum,
as we discuss in Section\ \ref{sec:conversion_to_km} below.

Standard errors are small, because of the high significance of the fits as expressed
by the F-ratio test.  
We present the best-fitting parameters for this IF, gate and channel range in Table\ \ref{tab:sample_parameters}.
These values are typical for our fits.
As we discuss in Sections\ \ref{sec:ceimf_size} and\ \ref{sec:systematic_effects} below,
errors in the estimated size are probably dominated by systematic errors.

\subsubsection{Fits to All Spectral and Gate Ranges}\label{sec:ceimf_size}

\begin{figure}[t]
\epsscale{.80}
\includegraphics*[width=0.98\textwidth]{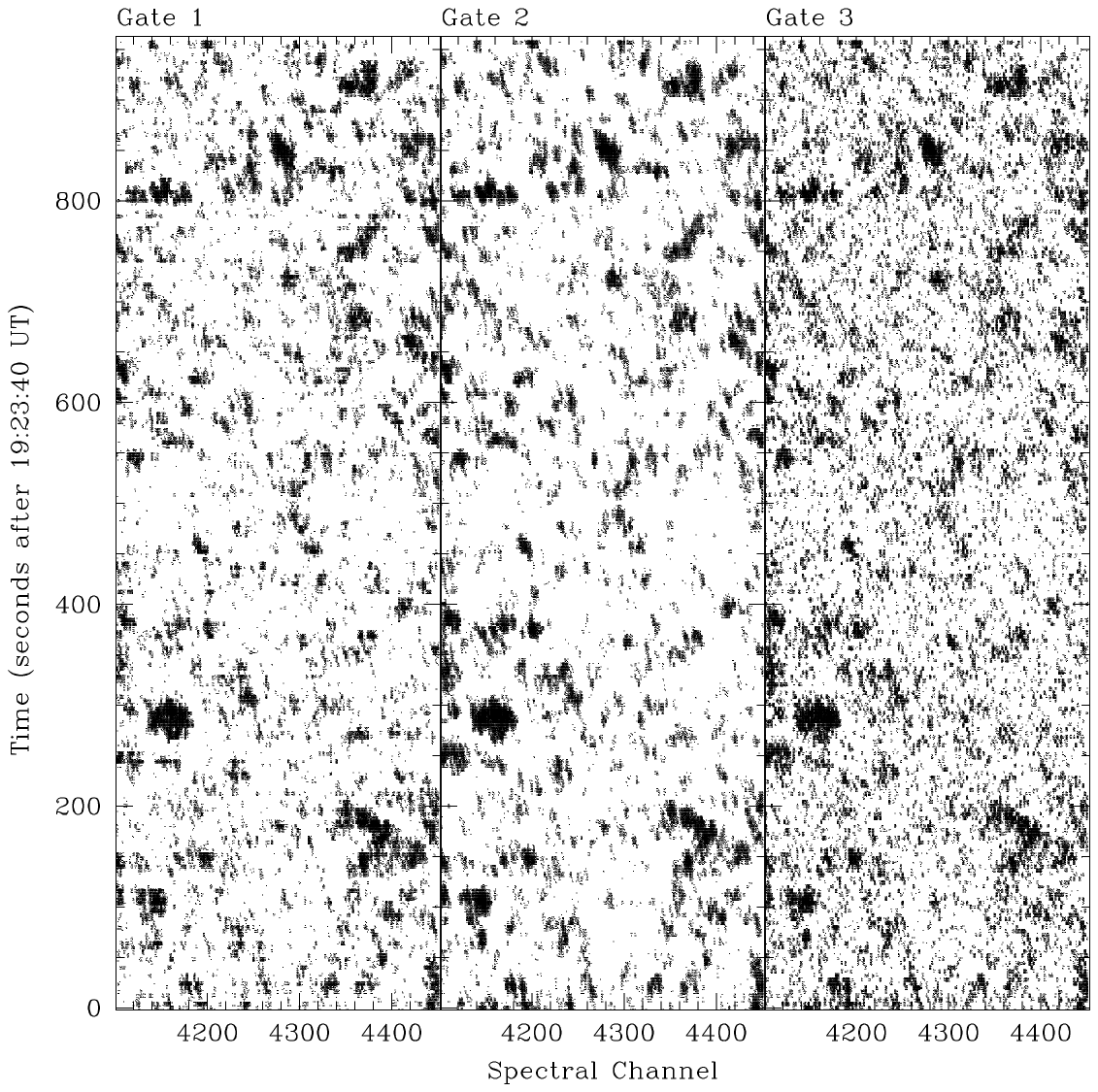}
\figcaption[]{
Dynamic spectrum, showing the real part of the visibility for a short frequency and time interval in 3 gates for IF1.
Amplitudes of the grayscales were equalized using the average real part in this spectral range, as shown in Figure \ref{fig:ampdraw}.
\label{fig:dynspec}} 
\end{figure}

\begin{figure}[t]
\epsscale{1.00}
\includegraphics*[width=0.98\textwidth]{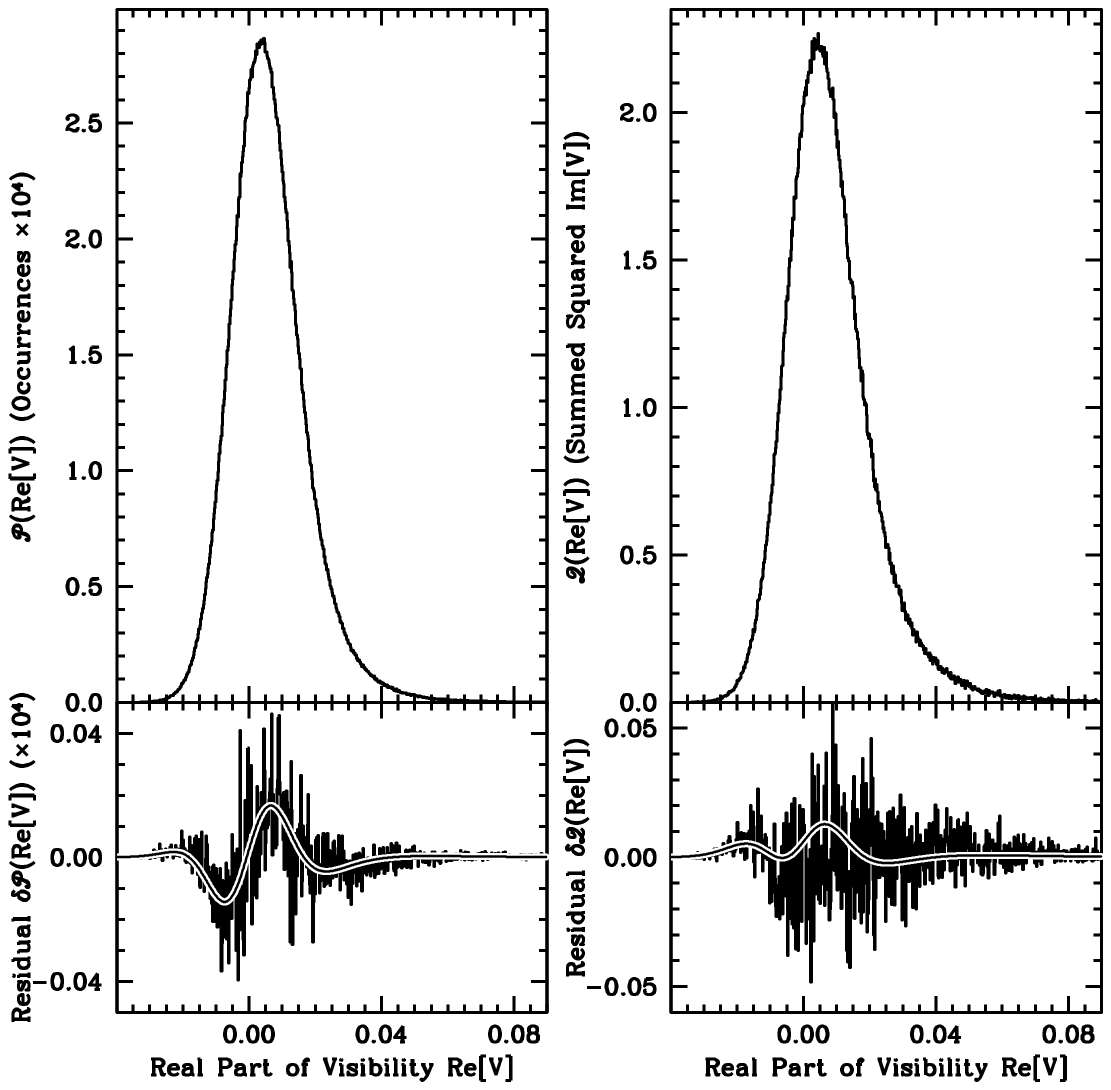}
\figcaption[]{Left panels: Observed distributions of visibility ${\mathcal P}$
for IF1, gate 1, channels 4096 to 5120. 
Upper: Distribution. Lower: Residuals to best-fitting model with zero size for pulsar.
Curve shows best-fitting model with finite size.
Right panels: Observed distributions of mean squared imaginary part ${\mathcal Q}$
for the same data. Upper: Distribution. Lower: Residuals to best-fitting zero-size model, and curve for difference of finite-size model from zero-size model.
\label{fig:observed_distributions}} 
\end{figure}

\begin{figure}[t]
\epsscale{1.00}
\includegraphics*[width=0.98\textwidth]{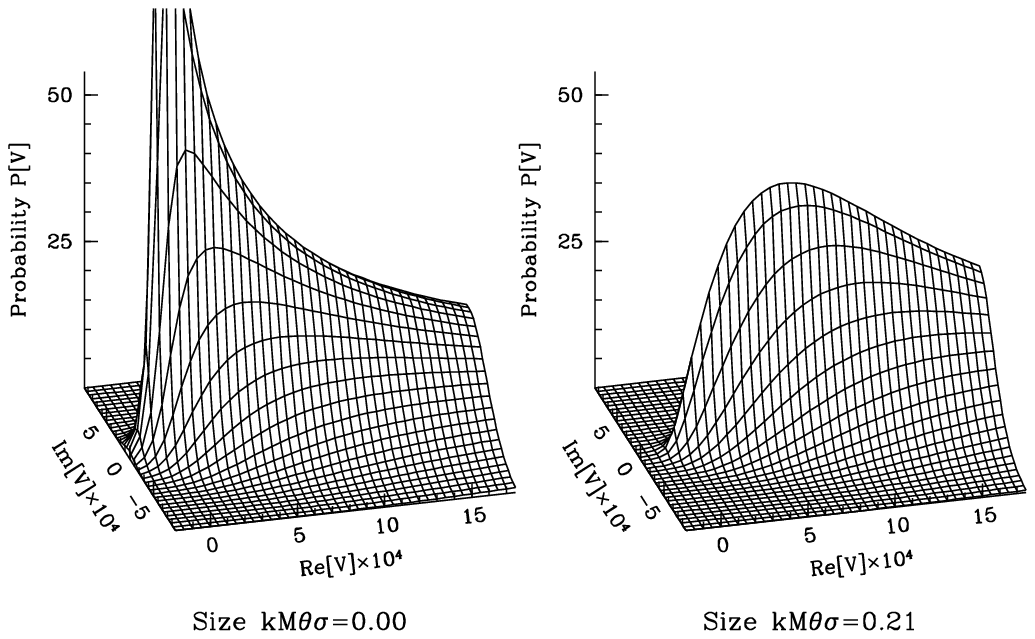}
\figcaption[]{Model distributions of visibility of a scintillating source in the absence of noise, 
for a pointlike source (left) and for a source of finite size (right).
Distributions assume amplitude $\kappa_0 = 0.006$ and correlation $\rho = 0.986$.  
\label{fig:size_nosize_nonoise_distribution}} 
\end{figure}

\begin{figure}[t]
\epsscale{1.00}
\includegraphics*[width=0.98\textwidth]{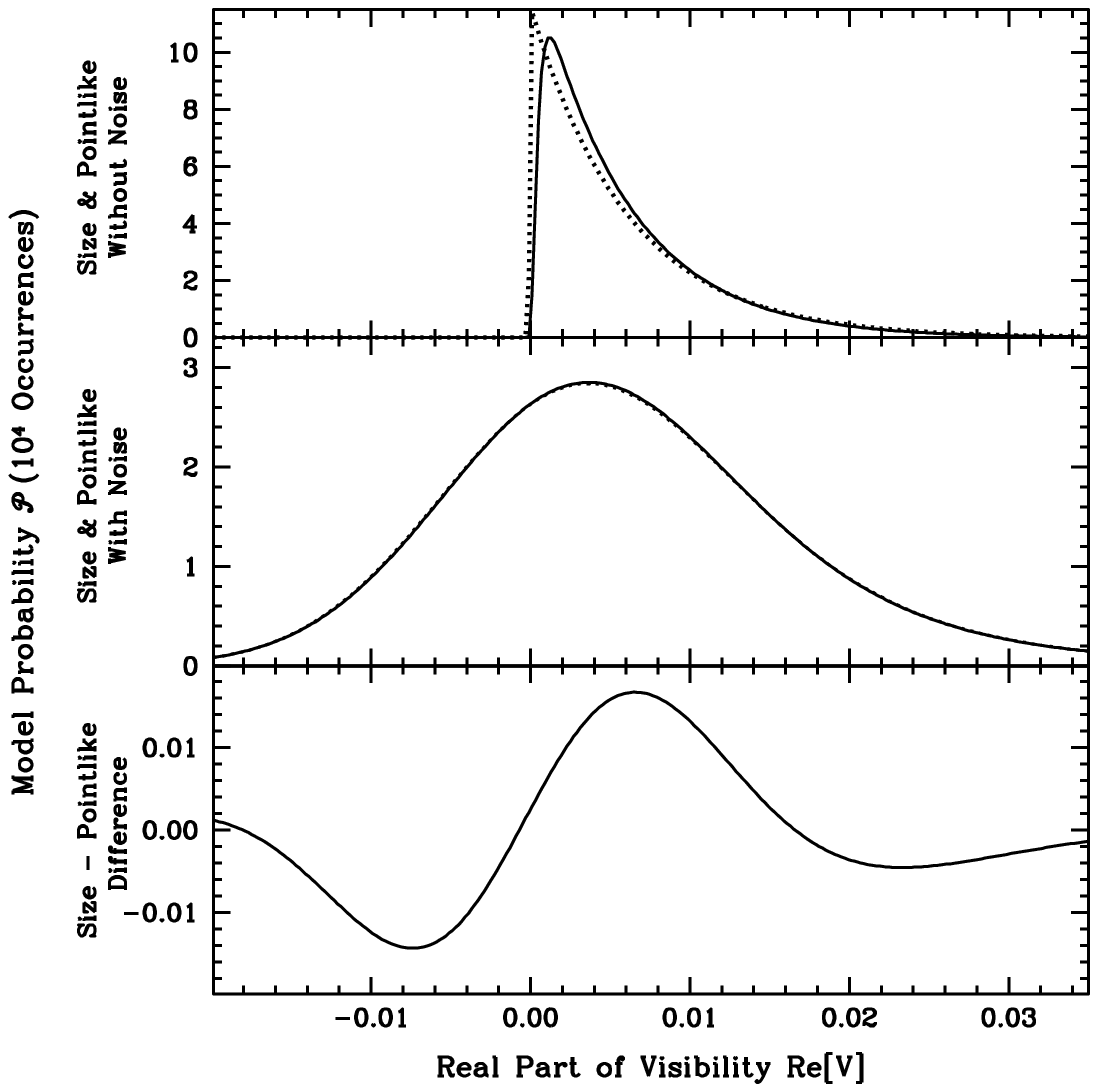}
\figcaption[]{Comparison of model distributions ${\mathcal P}$
for IF1, gate 1, channels 4096-5120.
Upper panel:
Projections for model without noise.
Dotted curve shows projection for pointlike source ($k M\theta\sigma =0$), solid curve for finite-size source ($k M\theta\sigma =0.21$).
Middle:
Projections including noise. Addition of the effects of noise 
makes the model curves nearly indistinguishable. 
Lower:
Difference of model curve for finite size from zero size, including noise.
The difference shows that the model with finite size has lower probability at negative ${\rm Re}[V]$ and larger positive 
${\rm Re}[V]$, but larger probability at intermediate ${\rm Re}[V]$.
\label{fig:projected_model_distributions}}  
\end{figure}

We fit our model to spectral ranges of 1024 channels in each of the 5 gates, for both IFs.
Our fits indicate that the pulsar has a rather large size at the start of the pulse, 
decreases in size over the first half of the pulse, and then increases in size again.
Figure\ \ref{fig:pulseshape_size} shows the results of fits, giving the scaled size $(k M \theta\sigma)$ and mean amplitude 
$(\kappa_0 + 2 \kappa_1)$ as a function of pulse phase.
Note that linear diameter is proportional to 
the square root of the fitted size parameter, $(k M \theta\sigma)^2$.
All of the fits are independent; each point represents an independent sample of the pulsar's emission,
fit completely independently using a priori parameters from independent grid searches (Section\ \ref{sec:gridsearch}).
The two IFs are shown by different symbols (crosses for IF1 and circles for IF2), which represent distinct frequency ranges
and thus completely different scintillation patterns.
At some phases, points from different gates overlap; at these points, a higher frequency range in one gate coincides with a lower 
frequency range in the previous gate.
Quite often the noise parameters for these different samples are quite different;
nevertheless, size and amplitude track one another with pulse phase, despite
the disparate origin of the samples.
Some of the fits do sample identical samples of the diffraction pattern from interstellar scattering, as Figure\ \ref{fig:dynspec} suggests:
however, these often lead to completely different results, indicating that the results arise from the pulsar, rather than from scattering.
For example, data that includes the 3 panels shown in Figure 1 yields scaled sizes of $(k M \theta\sigma) = 0.08$, 0.09, and 0.40.
(Of course, the grayscale in Figure\ \ref{fig:dynspec} emphasizes peaks
of visibility, whereas the most critical information on source size is found near zero visibility, as Figure\ \ref{fig:projected_model_distributions} shows.)
The dependence of fitted size on pulse phase, rather than on the frequency or details of the scintillation pattern, suggests that the
effect arises from the pulsar rather than the scintillation pattern.

Our sample comprises 60 independent fits, over the two IFs, 5 gates, and 6 spectral ranges.
A fit with finite size yielded significantly smaller residual for most of these:
for the 43 fits with fitted size parameters of $(k M\theta \sigma)^2 > 0.017$,
the reduction of the residuals was greater than 3\%.
This is significant at better than the 5-$\sigma$ level according to the F-ratio test.
Unsurprisingly, gate and spectral ranges with small fitted sizes,
and regions with small amplitudes at the beginning and end of the pulse, show the least significant results.
We do not display statistical errors from the fits on this figure; they are smaller than the plotting symbols. 

The scatter of independent measurements at a given pulse phase provides a measure of the precision of our results.
The residuals are dominated by variations among nearby bins, consistent with Poisson noise; 
whereas the models predict rather more slowly-varying differences across the histograms for ${\mathcal P}$ and ${\mathcal Q}$,
as Figure\ \ref{fig:projected_model_distributions} would suggest.
We made the bins narrow so as to follow the rapid variation of the histograms with ${\rm Re}[V]$,
as shown in Figure\ \ref{fig:observed_distributions}; we cannot broaden the bins without destroying this important information
by averaging.
We discuss averaging of the post-fit residuals over bins to visualize effects of the model in Section\ \ref{sec:pox} below.

\subsubsection{Conversion to km}\label{sec:conversion_to_km}

The angular width of the scattering disk $\theta$ 
is important for the conversion of the fitted size parameter $(k M \theta\sigma)^2$ to
$\sigma$ in km.
The angular width appears directly in this expression, and also affects the inferred magnification $M$ as we discuss below,
so that the linear size of the emission region $\sigma$ depends strongly upon $\theta$.
Unfortunately, our data do not provide a good 
determination of $\rho$,
as discussed in Section\ \ref{sec:rho} above,
and so do not allow accurate determination of $\theta$.
Consequently,
we adopt a simple conversion based on previous work,
while noting that more accurate measurements of $\theta$ and $M$
may revise the inferred conversion of $(k M \theta\sigma)$ to $\sigma$, in kilometers.

\citet{Gwi97} found angular broadening of $(3.3 \times 2.2)\  {\rm mas}$
(full width at half-maximum intensity), with the major axis at position angle $92^{\circ}$,
in observations at wavelength $\lambda=13\ {\rm cm}$.
We adopt this value here.
For simplicity in converting the observed $(k M \theta\sigma)$ to size $\sigma$, 
we assume a circular scattering disk with full width at half maximum of $\theta= 5.2\ {\rm mas}$;
this is the mean square of the major and minor axes.
For the scintillation bandwidth, we adopt the value of $\Delta \nu_s = 15$\ kHz from 
Section\ \ref{sec:time_freq_averaging} below.
With a distance to the pulsar of $D=287\ {\rm pc}$, as measured from the parallax of the pulsar \citep{Dod03},
the combination of angular broadening and scintillation bandwidth yields a magnification of $M=D/R = 3.1$ \citep{gwi93}. 
Consequently, the standard deviation of the Gaussian model for the source is
$\sigma_{\theta} = (k M \theta\sigma)\times 859\ {\rm km}$.
In this expression, the subscript $\theta$ serves
to indicate the dependence of the inferred source diameter on the angular size of the scattering disk,
and the magnification.
The full width at half-maximum diameter of the fitted model is $\sqrt{8 \ln 2} \sigma_{\theta}$.
The right-hand vertical axis on the lower panel in Figure\ \ref{fig:pulseshape_size} reflects these conversion factors.
The size is as large as hundreds of km, expressed as standard deviation of a circular Gaussian distribution.

Changes in the assumed parameters for angular broadening will modify the inferred emission size.
The angular scale of the scattering disk $\theta$ sets the scale of the scintillation pattern at the source, 
and thus the resolution of the lens; 
and $\theta$ sets the location of the scattering material along the line of sight, and so the magnification factor $M$. 
If the value assumed for $\theta$ changes, the inferred size $\sigma$ changes, even though the fitted parameter  $(k M \theta\sigma )$ remains unchanged.  
The net resulting effect is approximately proportional.
Thus, halving the assumed angular broadening reduces the inferred size expressed in kilometers by a factor of 2,
and so on.
The previous measurement of $\theta$ by \citet{Gwi97} covered an incomplete arc in the $(u,v)$ plane,
and depended on simpler approximations to the distribution of visibility for a scintillating source than the models used here;
the measurement should be repeated.
Refractive scintillation can change the angular broadening with time,
but the variation is expected to be only about 85 microarcseconds
for this line of sight \citep{Rom86,Nar92}.
If scattering material is distributed along the line of sight rather than in a thin screen, and some of it is located close to the source,
it can increase the magnification $M$ without a change in the measured angular broadening.
This would require scattering 
close to the pulsar, within the Vela supernova remnant, where density is expected to be low.

\subsubsection{Binned Residuals}\label{sec:pox}

We visualize the significance of our fits, 
by binning our residuals. This reduces effects of small-scale noise while allowing us to present the
signatures of source size in the data.
Philosophically, re-binning the residuals is similar to subtracting a model from the data.
For example, we subtract a model for source position, baseline length, and Earth orientation from the phases in the correlator,
so that the ``sky'' fringe rate can be reduced enough for time integration to recover the fringe
with high signal-to-noise ratio.
Fringing reveals the inaccuracies of the subtracted model.
Here, analogously, we remove a model for the distribution of visibility from the histograms of data,
reducing variations between neighboring bins from noise.
We then average the residuals of nearby bins so that the effects of source structure are more easily apparent.
This averaging reveals the degree to which our model fits match the observations.

Figure\ \ref{fig:poxres} shows the result of binning residuals,
again for the data in IF1, Gate 1, channels 4096 to 5120.
After fitting zero-size and finite-size models, the residuals were binned and averaged in groups of 20, for resulting widths of $20\, w = 0.0040$ in ${\rm Re}[V]$.
Averaging (rather than summing) keeps the normalization the same.
The dotted histogram shows the residuals from the fit with zero size for the source.
We also display the differences between the model for finite source size and for zero size as solid lines.
The curves are the same as those shown in Figures\  and\ \ref{fig:observed_distributions} and \ref{fig:projected_model_distributions}.
A zero-size source would show a flat, dotted histogram.
A model that perfectly corrected the deficiencies of the zero-size model
would track the dotted histogram perfectly, allowing for the finite widths of the bins.
Clearly, adding a size parameter explains most of the slowly-varying residuals.
Quantitatively, after binning the mean square residual is reduced by a factor of 13, by including a size parameter.
This is much larger than the reduction observed before binning.
As Figure\ \ref{fig:observed_distributions} suggests, Poisson noise
dominates the residuals without binning, and is greatly reduced by binning.

The model fit is imperfect. Some systematic variations remain,
more so in some ranges of pulse phase than others.
Figure \ref{fig:triplepox} shows residuals for the zero-size model,
and differences between finite-size and zero-size models, for representative spectral ranges in three gates.
Fits to the same spectral range and same gate, but different IF,
resemble one another strongly, as expected because they lead to nearly the same fitted size.
Noise and amplitude parameters vary between IFs, and even more between gates and channel ranges,
so that the binned residuals are not identical.

The upper pair of panels of Figure \ref{fig:triplepox} show IF2, gate 1, channels 4096 to 5120: these data are equivalent to 
those in Figure\ \ref{fig:poxres} in pulse phase and spectral range, but are for the other IF. 
The data are thus completely independent.
Introduction of a size parameter reduces the mean square residual of $\mathcal P$ by 92\%, and $\mathcal Q$ by 12\%.
The best-fitting size parameter is $\kappa_1/\kappa_0 = (k M \theta \sigma)^2 = 0.043$,
or scaled size $(k M \theta \sigma) = 0.21$.
The IFs agree closely in residuals and in fitted size.
This range lies on the rising part of the pulse profile, before the peak.

The middle pair of panels show IF1, gate 2, channels 2048 to 3072.
This range lies near the peak of the pulse.
The best-fitting size parameter is $\kappa_1/\kappa_0 = (k M \theta \sigma)^2 = 0.006$,
among the smallest results we obtain.
Before binning, introduction of the size parameter reduced the Poisson-weighted residuals by less than 1\%;
this reduction is significant at the 2-$\sigma$ level according to the F-test.
After binning, the reduction in mean square residual for ${\mathcal P}$ is 32\%;
however,  zero-size
and finite-size models both fit $\mathcal Q$ almost equally well.
Some systematic effects appear, particularly the sharp downward spike 
in $\mathcal P$ near ${\rm Re}[V]=0$.

The lower pair of panels show IF2, gate 3, channels 4096 to 5120.
This range lies on the trailing side of the pulse, near where it briefly flattens into a plateau.
At shorter observing wavelength, this plateau becomes a second component that arises close to this pulse phase.
Introduction of a size parameter decreases the residuals,  by 10\% in $\mathcal P$
and by 4\% in $\mathcal Q$ before binning, and by 54\% and 24\% after binning.
Again the influence of $\mathcal Q$ on the overall fit is small, because zero-size
and finite-size models both fit $\mathcal Q$ well.
The best-fitting size parameter is $(k M \theta \sigma)^2 = 0.039$,
or $(k M \theta \sigma) = 0.20$.
Although the fit to $\mathcal P$ is good,
some systematic variation appears as a shifting of the histogram relative to the best-fitting model;
our circular-Gaussian model does not accommodate such a shift.
We are investigating a variety of simple models for the systematic variations,
as we discuss briefly in Section\ \ref{sec:alternative} below.

\subsection{Systematic Effects}\label{sec:systematic_effects}

The effects of size that we observe in our data include its
W-shaped signature in the distribution of visibility ${\mathcal P}$,
its variation with pulsar phase,
and its constancy for different spectral ranges and samples of the scintillation pattern at the same pulse phase.
The W-shaped
signature of finite size on the distribution of visibility is quite characteristic, as 
Figures\ \ref{fig:observed_distributions},\ \ref{fig:projected_model_distributions}, and \ref{fig:poxres} suggest,
and appears for both IFs and many pulse gates and spectral ranges.
The agreement of fitted size is good between IFs, for overlapping spectral ranges at different frequency but the same pulse phase,
and between nearby but different spectral ranges.
Effects that can match all of these, or even only the latter two, are nearly all associated with geometrical effects at the pulsar.

\subsubsection{Errors in Model Parameters}

Further significant changes in the magnification factor from changes in pulsar distance seem unlikely.
The distance from parallax is far more reliable than that from dispersion measure.
In principle a model that included significant scattering in the immediate neighborhood of the pulsar, as well as scattering in the surrounding Vela
supernova remnant, could increase the effective magnification by bringing the lens closer to the pulsar.
Temporal broadening at low frequencies suggests that scattering material is concentrated into a thin screen \citep{Joh12c}.
Previous measurements of the angular broadening $\theta$ should be revisited, as described in Section\ \ref{sec:conversion_to_km} above.
Further studies of the angular broadening, using long baselines with different orientations,
can help to improve on the earlier results.
This may change the inferred linear size $\sigma$.

\subsubsection{Instrumental Effects}\label{sec:instrumental_effects}
	
Instrumental effects are unlikely to reproduce the signature of source size on the distribution of visibility,
as shown in Figure\ \ref{fig:poxres}, because they tend to vary with pulse gate and frequency within the passband.
For example, for both IFs, the noise parameter $b_0$ shows a similar pattern for each pulse gate, 
reflecting variations in the gain and noise in the passband, with an overall offset for each gate that
reflects changes in quantization noise \citep[see][Section 5.1]{VelaNoise}.
Effects of quantization, in particular, can be expressed in the spectral domain as a gain and a change in noise,
and so are absorbed into the amplitude parameter and $b_0$.
Effects of variation of noise parameters would be expected to appear as differences between overlapping gates.

The direction of linear polarization varies smoothly across the pulse.
We observe only left-circular polarization, the fraction of which varies little.
Scintillation in the interstellar medium is nearly polarization-independent \citep{Min96,Spa01}.
Defects in separation of polarizations at the antennas could produce artifacts that vary with pulse phase,
but these can be represented accurately as pulse-phase-dependent gains.
They would not produce the distortion of the distribution of visibilities that we observe as the signature of source size.

Saturation effects can be important at large antennas such as Tidbinbilla.
For our observations, these might affect particularly strong pulses, such as the giant pulses reported by \citet{Joh01}.
However, such pulses carry an insignificant fraction of the total flux density of the pulsar. 
Saturation effects can also reduce visibility for the very strongest scintillations, at least for baselines involving two large antennas 
\citep[][Section 4.5]{Gwi00}. These effects are expected to distort the distribution of visibility at the largest values,
but are not likely to reproduce the W-shaped signature of finite size we observe.

\subsubsection{Calibration and Fringing Effects}

We perform relatively little calibration on the data.  We do not change amplitude, particularly within a scan or with frequency.
We do change phase, in fringing.  
Mis-fringing is possible for observations of a scintillating pulsar in the speckle limit, because
the phase and amplitude vary on the frequency and time scales of scintillation \citep{des92},
so that the model used for fringing can be challenging to fit.
Mis-fringing would tend to increase phase variations, while keeping amplitude the same.
Such effects are reduced for a short baseline, such as the Mopra-Tidbinbilla baseline, because the phase variation is less.
If the phase errors were very large, it could potentially alter the distribution of visibilities in our projections, 
by altering the relative sizes of real and imaginary part.
We reduced errors from mis-fringing by fitting our fringe model over approximately 2 decorrelation times and 
800 scintillation bandwidths.
The wide bandwidth reduces effects of scintillation-based phase variations.
The short time was chosen to match any possible ionosphere- or atmosphere-induced phase variations.

Effects of errors in the fringe model on size would be expected to reproduce among gates, 
since the fringe model from Gate 2 was applied to all gates (with the exception of a single phase across the band,
fitted for each gate);
however, size varies with pulse phase, rather than by channel range within gates,
and is near zero for some spectral ranges near the pulse peak.
For channel ranges with sufficient signal-to-noise ratio,
such as the later channels of Gate 1 and most channels of Gate 3,
we found that fringing to the data within that gate, rather than using the model from Gate 2, produced identical results.
Similarly, applying the model from other strong gates to Gate 2 yielded nearly the same results.

\subsubsection{Time and Frequency Averaging}\label{sec:time_freq_averaging}

Effects of the finite sampling of the diffraction pattern are small, and are expected to be constant across the pulse.
Sampling over finite intervals of time and frequency blurs together correlated samples of the diffraction pattern,
and so is indistinguishable from effects of source size \citep{Joh12a}.
For given scintillation timescale $t_{\rm ISS}$ and scintillation bandwidth $\Delta\nu$,
effects of sampling can be calculated analytically \citep[][Section 2.2]{Gwi00}.

We estimate the scintillation timescale and bandwidth from our observations on the Mopra-Tidbinbilla baseline,
using data near the pulse peak, IF2, Gate1, channels 2048 to 3072.  
We find correlation functions for the observed real part of interferometric visibility,
and fit expected functional forms to them \citep[see][Equations 5, 7]{Gwi00}.
Figure\ \ref{fig:scintillationscales} displays the data  and results.
We have normalized the correlation functions using the amplitude and offset of the best-fitting functional forms.
The data point at zero lag is omitted, because it is elevated by effects of noise and variations of the amplitude with time.
Both correlation functions extend to large lag, off the right of the displayed panels; these are included in the fits, but not displayed here.

A fit of the expected Gaussian function to the temporal correlation function yields a scintillation timescale of $t_{\rm ISS}= 8.99\pm 0.17$\ s.
Our sampling time of 2\ sec is then expected to decrease the mean square flux density by a factor of $f_{t} = 0.994$ \citep[][Equation 8]{Gwi00},
while leaving the average flux density unchanged.
The variation of $f_{t}$ from the standard error of the fit is less than 0.001.

A fit of the expected Lorentzian function to the frequency correlation function yields a scintillation bandwidth of $\Delta\nu = 15.2\pm 0.5$\ kHz.
Our channel bandwidth of 1.95\ kHz is then expected to decrease the mean square flux density by a factor of $f_{f} = 0.998$ \citep[][Equation 6]{Gwi00}, leaving the mean flux density unchanged.
The observed correlation function shows systematic departures from the best-fitting Lorentzian function,
with a sharper peak and elevated correlation at intermediate lag, with a peak near a lag of 70\ kHz.
A Kolmogorov rather than a square-law structure function,
and a departure of the scattering disk from isotropy, can produce such effects\ \citep{Gwi01,Joh12a}.
Temporal variation of the emission on short timescales will also introduce correlations in frequency\ \citep{Gwi11a}.
Models including such effects will fit better.
However, as the figure shows, the correlation falls to half its peak value at a lag of approximately 14\ kHz,
providing a clear characteristic scale.
The variation of $f_{f}$ from the standard error of $t_{\rm ISS}$ is less than 0.001.
Even with $\Delta\nu=12$\ kHz we still find $f_{f}=0.998$.

Together, the effects of finite sampling in time and in frequency will increase the fitted size.
Under the assumption that the source is pointlike, so that $(kM\theta\sigma)=0$,
and we misinterpret the effects of averaging in time and frequency as source size,
our sampling parameters lead to a fictitious size parameter of 
$(k M \theta \sigma) = 0.06$,
where we have related effects of finite sampling to size using the expressions for modulation index from averaging \citet[][Equation 9]{Gwi00}
and for source size \citet[][Equations 34, 35]{Gwi01}.
This value is indeed nearly the minimum that we observe, near the pulse peak, as Figure\ \ref{fig:pulseshape_size} shows.

The effects of finite sampling in time and frequency are constant with pulse phase, for a pointlike source.
This is clearly inconsistent with our results discussed in Section\ \ref{sec:results_discussion} above.
Variation of amplitude with time can affect estimates of the scales of scintillation:
variability on long timescales can affect the correlation with time, and on short timescales can affect the correlation in frequency \citep{Gwi11a,Gwi11b}.
Erroneous estimates will not produce variations of source size with pulse phase. 
Direct effects of source variability on estimates of source size are discussed elsewhere in this paper (Sections\ \ref{sec:amp_variations},\ \ref{sec:noiseandamplitudevariations}).

\subsection{Alternative Models}\label{sec:alternative}

Our fit assumed a circular-Gaussian distribution of emission at the source.
More precisely, since the scattering disk is presumably elongated, we have fit the data with a model for a source
with the same axial ratio as the scattering disk.
This description simplifies the mathematics.  
Some of the residuals in Figure\ \ref{fig:triplepox} show systematic residuals 
that suggest that a more complicated model might provide a better fit.
A source elongated along a single direction provides a model arguably as simple as the one we use;
the size inferred for such a fit is approximately $\sqrt{2}$ greater than the diameter for a circular Gaussian. 
A source of arbitrary axial ratio requires a more complicated model;  
we will discuss such models in conjunction with data on longer baselines.
Of course, an infinite set of models can be fit to any finite data set;
we choose the circular-Gaussian case as providing a simple parameterization that agrees well with our data.

Another simple model is a ``core-halo'' model, including a pointlike core superposed with an extended halo.
In the simplest case, this model would include a pointlike core and a halo so extended that it did not scintillate at all.
Such a model would produce the distribution of visibility for a scintillating point source, 
as shown for example in the left panel of 
Figure\ \ref{fig:size_nosize_nonoise_distribution}, offset to the right, by convolution with the delta-function distribution of visibility expected
for a non-scintillating source.
To match our observations, the two components of this distribution would have to 
change their relative magnitudes over the course of the pulse,
and to both disappear when the pulsar is ``off": we see neither in the empty Gate 6.
Moreover, the size of the ``halo'' would have to be large: much greater than 800\ km.
We regard this model as more complicated, and less natural, than a circular-Gaussian model.
Such a model appears to fit less well than a circular-Gaussian model at either side of the pulse maximum,
where size and amplitude are both relatively large, although it fits significantly better than a zero-size model at some ranges of pulse phase. 
	
\subsection{Comparison with $\lambda = 13$ cm results}

\begin{figure}[t]
\epsscale{.80}
\includegraphics*[width=0.98\textwidth]{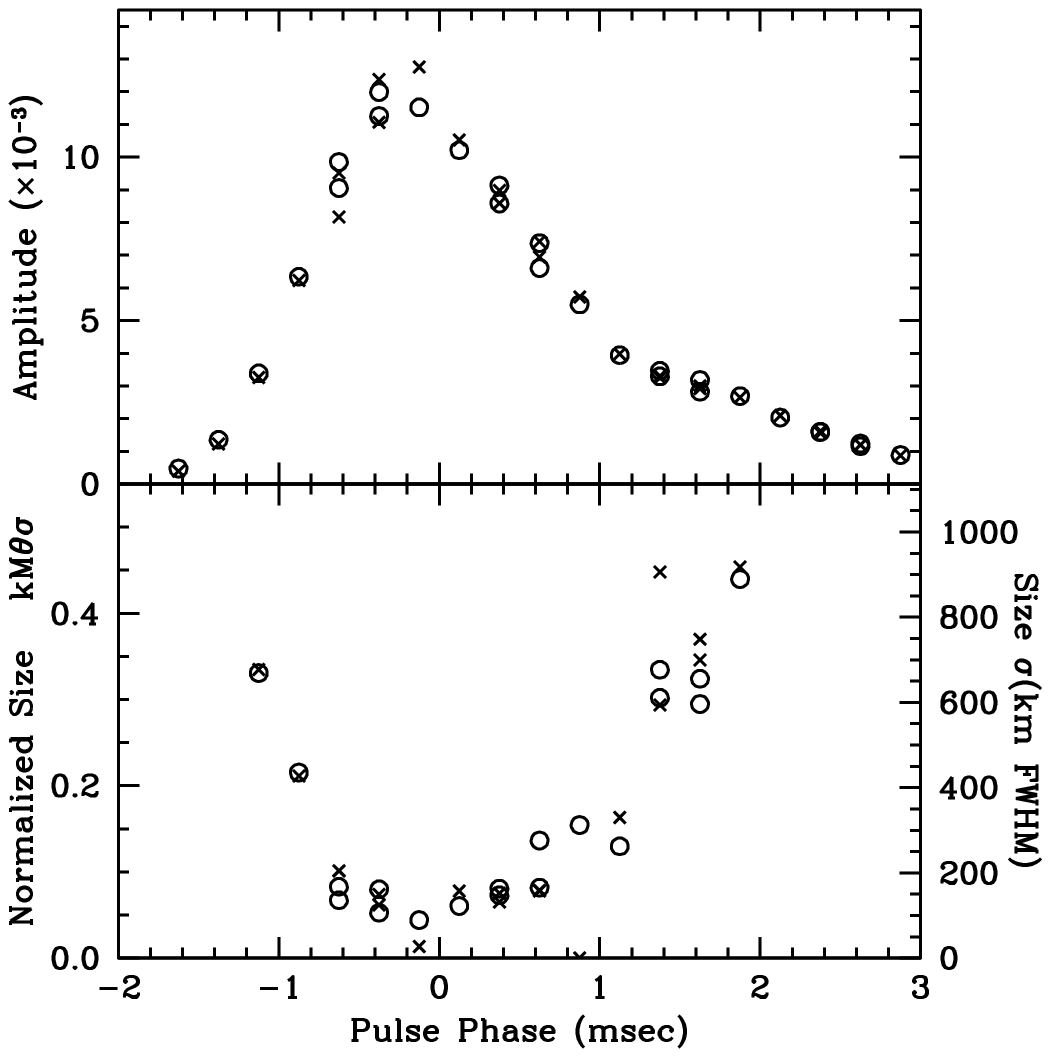}
\figcaption[]{Best-fitting amplitude $\kappa_0+2\kappa_1$ (top panel) and normalized source size $(k M \theta\sigma)$ (lower panel) plotted with pulse gate,
for 4 gates in 6 spectral ranges. 
The model for the emission region assumes a circular Gaussian distribution of emission.
Crosses show IF1, circles show IF2.
Right-hand axis on lower panel shows full width at half maximum of size of emission region in km, estimated as described in Section\ \ref{sec:conversion_to_km}.
\label{fig:pulseshape_size}}   
\end{figure}

\begin{figure}[t]
\epsscale{.90}
\includegraphics*[width=0.98\textwidth]{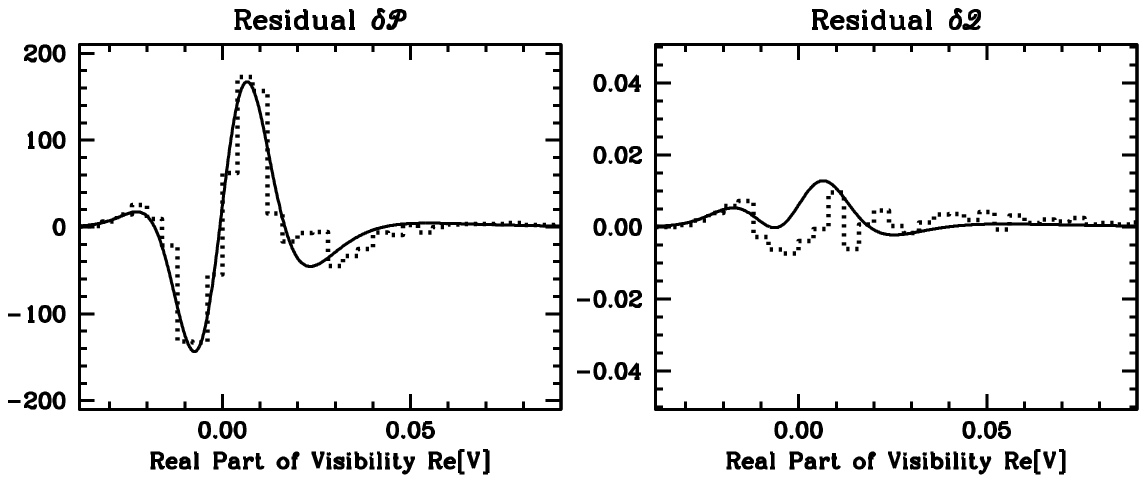}
\figcaption[]{
Plots of the binned differences of the data from the best-fit zero-size model (dotted histogram),
and of the best-fitting finite-size model from that zero-size model (solid curve).
Agreement of the solid curve with histogram shows that finite-size model can explain many of the binned residuals.
Data are for IF1, gate 1, channels 4096Ð5120, as shown in Figure 3.
The curves are identical to those shown in the lower panels of Figure 3, and the curve in the right panel is identical to that shown in the lower panel of Figure\ \ref{fig:projected_model_distributions}.
%
%
\label{fig:poxres}}   
\end{figure}

\begin{figure}[t]
\epsscale{.90}
\includegraphics*[width=0.98\textwidth]{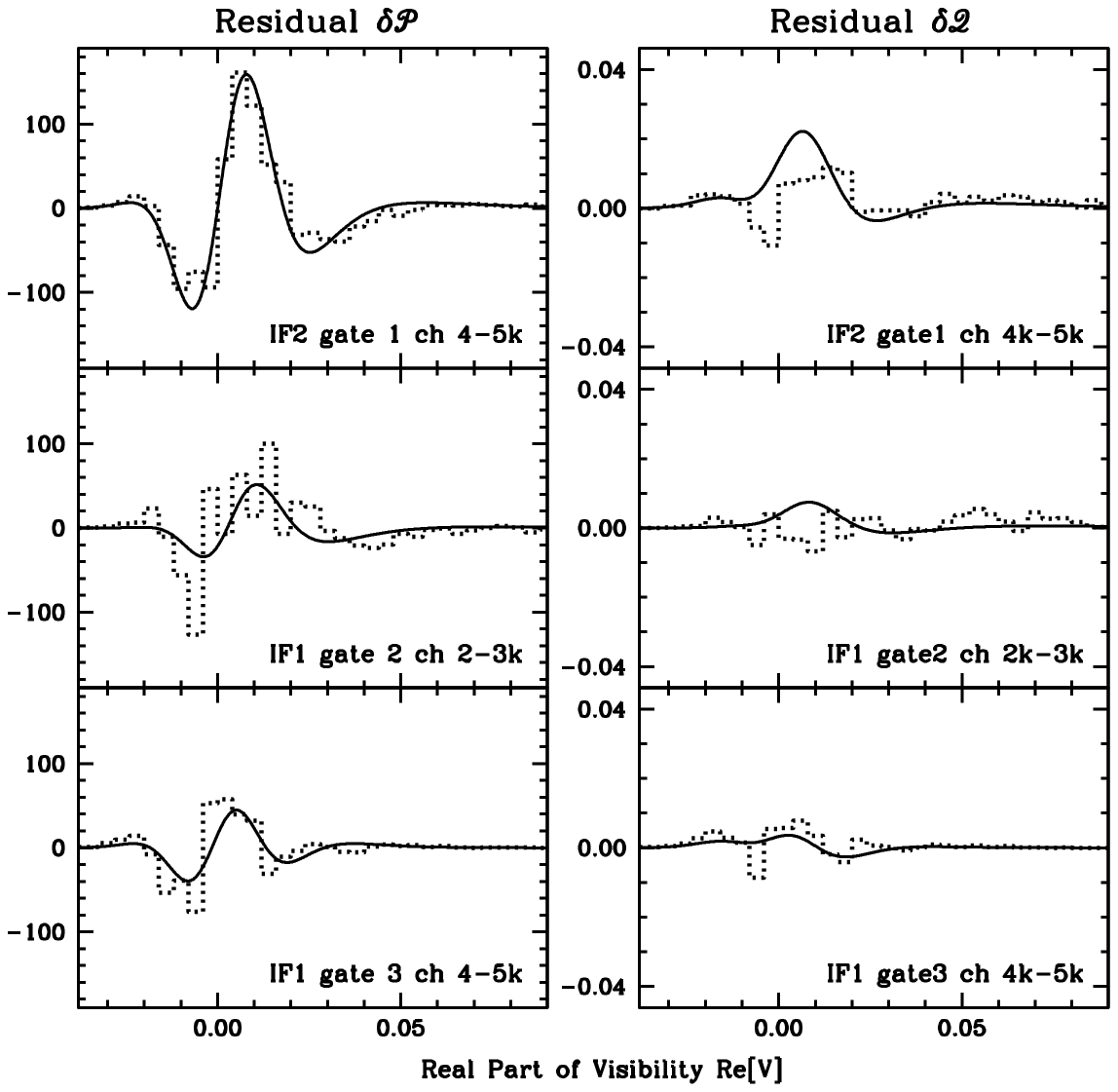}
\figcaption[]{Binned residuals in 3 gates for different pulsar gates and spectral ranges from that shown in Figure\ \ref{fig:poxres}.
Left panels show residual $\delta{\mathcal P}$, right panels show $\delta{\mathcal Q}$.
Legends in panels indicate IF, gate, and channel ranges.
The top panels show the same channel and gate range as Figure\ \ref{fig:poxres},
but in the other IF.
\label{fig:triplepox}}   
\end{figure}

\begin{figure}[t]
\epsscale{.90}
\includegraphics*[width=0.98\textwidth]{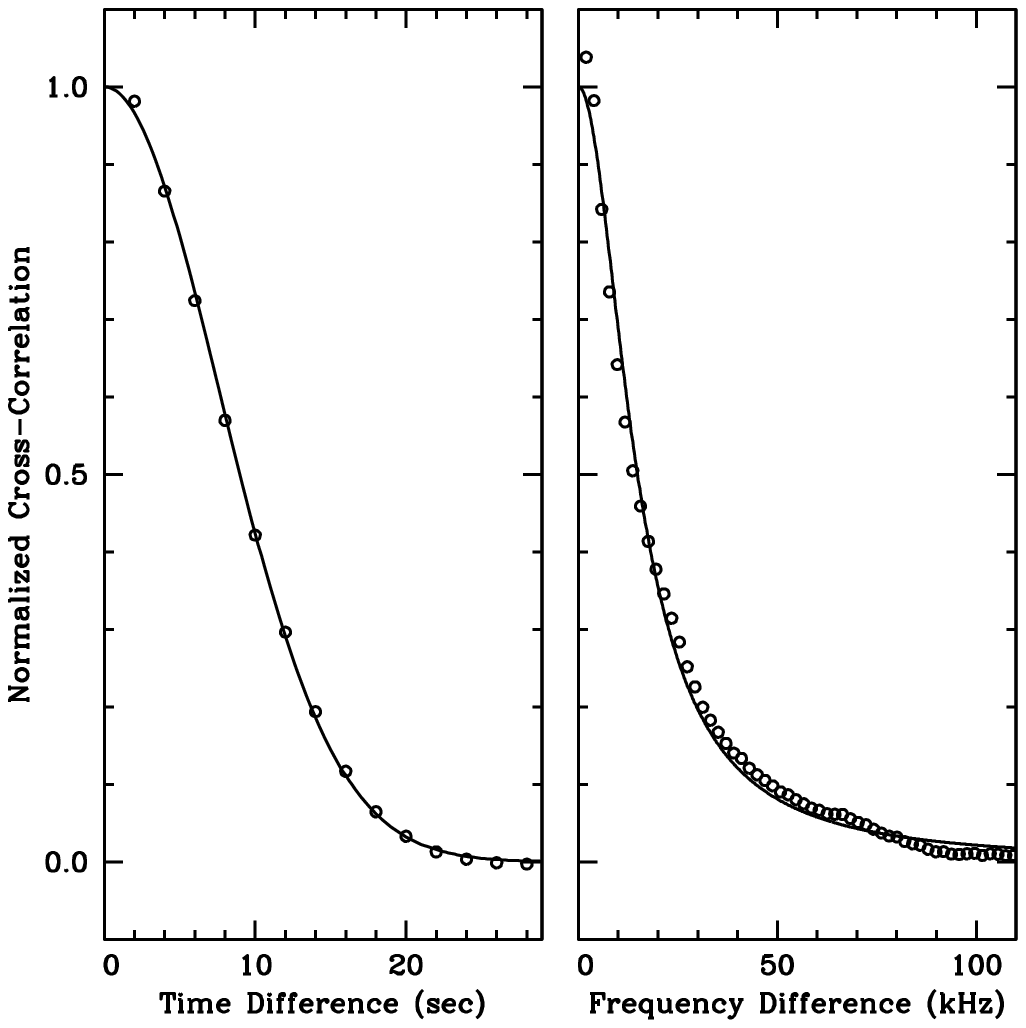}
\figcaption[]{Correlation function of real part of visibility in time (left panel) and in frequency (right panel),
with best-fitting theoretical forms for the short Mopra-Tidbinbilla baseline, near the peak of the pulse at
IF2, Gate 2, channels 2048 to 3072.
Correlation functions extend to large lags, not shown but included in the fits.
\label{fig:scintillationscales}}   
\end{figure}

We reported a size of the Vela pulsar's emission region of approximately 300 km, at 13\ cm observing wavelength, in previous work \citep{Gwi97,Gwi00}.
The three gates used there correspond roughly to ranges of pulse phase
of $-1.2$ to 0.2\  msec, 0.2 to 1.2\ msec, and 1.2 to 3.7 msec,
in Figure\ \ref{fig:pulseshape_size}.
We reported size parameters of 
$(k M \theta \sigma)^2$ of 
$0.091\pm 0.009$, 
$0.070\pm 0.009$,
and 
$ 0.020\pm 0.020$
in three gates across the pulse, at $\lambda=13\ {\rm cm}$ observing wavelength.
(Note: The exponent ``2'' 
was omitted in the legend for these quantities in the first line of the body of Table 4 of \citet{Gwi00}).
For comparison with the lower panel of Figure\ \ref{fig:pulseshape_size},
the square roots of these measured quantities should be multiplied by the ratio of wavelengths of the observations, 18/13.
The resulting sizes are comparable to those inferred here, except for the third gate,
for which the size at 13\ cm wavelength is significantly smaller.
The 13-cm gate extends well past the end of the pulse at 18 cm;
the pulse is wider at the shorter wavelength, and flux density in the gate is significantly higher,
because of the emergence of a second peak.  This profile difference might be responsible for the emission size discrepancies.

The observations and analysis at 13-cm wavelength differed in detail from those presented here.
One difference was that we assumed a circular source, rather than one with the same aspect ratio as the scattering disk;
this would tend to increase the size parameter $(k M \theta \sigma)^2$ estimated at 13\ cm wavelength.
We also used an exponential model; this is a reasonable approximation for the shorter baseline at shorter wavelength.
The revised distance to the pulsar from parallax measurement,
of  $D=287\ {\rm pc}$ \citep{Dod03} rather than $500\ {\rm pc}$,
changes the magnification factor $M$ from $1.5$ to $3.1$
and so decreases the estimated size of the pulsar.
The integration time was longer and the spectral-channel bandwidth was higher,
as allowed by the increased decorrelation scales of the scintillation.
However, the channel bandwidth was great enough that the effects of averaging in frequency were non-negligible 
\citep[see][Section 2.2]{Gwi00}.
There were also fewer scintillation elements sampled.
The model was fit to amplitude rather than to interferometric visibility;
consequently, the effect of size on the distribution function is less simple.
The model for self-noise omitted the quadratic term, although 
this omission was justified because the effects of finite source size appear at small amplitude,
where the quadratic term is small.  A 2-dimensional distribution of noise
as given by $b_0$ and $b_1$ is critical for an accurate fit to the distribution of amplitude, however.

\subsection{Comparison with Pulsar Geometry}\label{sec:geometry}

The pulse profiles of most pulsars can be divided into ``core'' and ``cone'' components, after the inferred shape of the beam \citep{ran83,ran90}. 
Vela is usually classified as a ``core single" pulsar \citep{ran93}, as might be expected from its young age and short period;
its narrow pulse, with width nearly constant with observing wavelength;
and its strong polarization at meter wavelengths.
A second component emerges at observing wavelength $\lambda \sim 10$\ cm, 
although it is not clear that this represents a subsidiary peak \citep{Ker00}.  
Vela fits the scaling relation of pulse width with period for conal pulsars well \citep{ran93}.
This scaling matches the opening angle for a dipole field near the star's surface, suggesting that 
core-single emission arises from near the star's surface.
The radio beam is not well aligned with the magnetic axes suggested by high-energy emission,
as is not uncommon for single-core components, raising the possibility of other geometry for this type of emission.

Unlike most core-single pulsars, Vela shows an organized pattern of polarization;
indeed, it was the archetype of the rotating-vector model,
which provides the fundamental pattern for conal emission \citep{rad69,kom70}.
The angle between the magnetic axis and the line of sight is $6^{\circ}$ in this model \citep{Joh01};
if one assumes that emission arises where the last open magnetic field lines are tangent to the line of sight,
and that the rotation axis lies in the plane of the sky,
the altitude of emission is 200\ km. 
The inferred altitude is significantly 
greater if the rotation axis is not in the plane of the sky.

\citet{kri83} analyzed the polarization properties of the Vela pulsar, in bins in amplitude of individual pulses.
They used the mapping of polarization onto dipole field geometry, and the difference in arrival time,
to infer the location of four distinct pulse components.
They found that earlier emission components arise from higher above the neutron-star surface than later components,
with a spread in altitudes of 500\ km.

In contrast, our measurements are sensitive to the lateral extent of the emission region, rather than its altitude.
These two are related by the height of the emission region and, plausibly, the geometry of the magnetic field.
Detailed discussion is beyond the scope of this paper,
but we note that the inferred lateral sizes tend to suggest emission altitudes at least as great as those inferred from field geometry 
\citep[][and references therein]{Dyk04,Gan05,Joh12b}.

The observed size might result from refraction or scattering in, or near, the pulsar's emission region.
The last closed field lines provide zones where plasma waves might grow, and then
then propagate along the field until converted to electromagnetic waves \citep{Bar86}.
Such an emission process might preserve much of the geometry and temporal variations of the emission,
while translating it to a higher altitude \citep{hir01}.
Interestingly, polarization appears to be perpendicular to the curvature of the magnetic field lines for the Vela pulsar,
which suggests reprocessing after emission \citep{Lai01}.  
Individual samples of the electric field of the pulsar show log-normal statistics, suggestive of multiple, random 
contributions to amplification \citep{cai03}.
Radiation emitted at low altitudes may also be refracted or scattered at higher altitude \citep{lyu00}. 
Reprocessing might include interaction with X-ray and gamma-ray emission \citep{har08,pet09}. 
Thus, a variety of evidence suggests that the site of the initial emission may be distinct from the location at which the radiation is released
to propagate, with scattering in the interstellar medium, to the Earth.

\section{SUMMARY}\label{sec:summary}

We have described analysis of the distribution of visibility of the Vela pulsar at $\lambda=18$\ cm wavelength.
We observed the pulsar in two IFs of 16\ MHz bandwidth each.  We observed the pulsar in 5 gates across the pulse, each 1\ msec wide.
The signal was not dedispersed, so higher frequencies sampled later pulse phase than lower frequencies.
We fringed the data and from it formed distributions of visibility projected into bins along the real axis ${\mathcal P}\left( {\rm Re}[V]\right)$ and the
second moment of imaginary part in each bin ${\mathcal Q}\left( {\rm Re}[V]\right)$.  

We calculated a theoretical model for the distribution of visibility.  This model used the distribution of interferometric visibility for a scintillating source
\citep{Gwi01}. 
We calculated models for a source of zero size, and for a source of small but nonzero size. We assumed a Gaussian distribution of flux density with the
same axial ratio as the scattering disk.  We then added the effects of noise and self-noise using convolution with a non-stationary kernel.
We modeled noise with a second-degree polynomial in phase with source visibility, and the linear terms of that polynomial at quadrature,
as predicted by theoretical models for noise and exhibited by our data \citep{VelaNoise}.
We also included a parameter for effects of intrinsic amplitude variability of the source, on timescales longer than the 2-sec integration time.
We find that effects of size appear as a W-shaped difference of the best finite-size model from the best zero-size model,
as a consequence of the shape of the underlying distributions of visibility.

We fit our models to the observed distributions using the Levenberg-Marquardt method.
We weighted residuals by Poisson statistics for ${\mathcal P}$ and by statistics appropriate for mean square elements 
of a Gaussian distribution for ${\mathcal Q}$, with constant weight below a cutoff of $N\leq 100$ samples in a bin. 
We searched parameter space using a grid search with the exponential $\rho=1$ form for the visibility distribution,
and found that the variations of the summed, squared weighted residuals was quite simple.
We used the minimum parameters from this search as initial parameters for Levenberg-Marquardt iterative fits.

Residuals to our fits for zero size show the characteristic W-shaped signature of finite size.  Independent fits for finite size describe the data well,
matching the data and presenting the expected form when differenced with the zero-size model.
We find that the resulting inferred, intrinsic size of the pulsar emission region
first decreases across the pulse, then increases.  The maximum size is approximately 800\ km (FWHM of our Gaussian distribution), and the minimum
is near zero.
The reduction in residuals ranges up to 10\%.
The statistical significance of including the size parameter is high, 40-$\sigma$ or more in some cases.
Residuals are dominated by Poisson noise,
from the finite number of samples in each bin.  

To help visualize effects of the fits, we binned the residuals after fitting.
Binning reduces Poisson noise, and removes the relatively rapid variation of the distribution,
leaving the slower variations from source size.
The results show that the distribution of visibility matches the W-shaped signature of finite source size well,
as we observe in a test case even before binning.
After removing most Poisson noise by binning, 
introduction of the fitted size parameter reduces residuals by as little at 30\% and as much as 92\%.
Some gates show evidence for 
residual systematic differences of the distribution from the model.
We consider various systematic effects that could modify our results, and compare results with previous observations at 
$\lambda=13$\ cm observing wavelength.
We briefly compare our results with previous work on geometry of pulsar emission.

\acknowledgments

I gratefully acknowledge the VSOP Project, which is led by the Japanese 
Institute of Space and Astronautical Science in cooperation with many 
organizations and radio telescopes around the world.
I am grateful 
to the DRAO for supporting this work with extensive
correlator time.
We thank the U.S. National Science Foundation for financial support for this work
(AST 97-31584 and AST-1008865).


\clearpage

\clearpage
\begin{deluxetable}{llll}
\tablenum{1}
\tablecolumns{4}
\tablewidth{0pc}
\tablecaption{Best-Fit Parameters: IF 1, Gate 1, Channels 4096-5120}
\tablehead{
\multicolumn{2}{c}{Parameter}&
\multicolumn{1}{c}{Fitted Size}&
\multicolumn{1}{c}{Zero Size}\\
\multicolumn{1}{c}{Type}&
\multicolumn{1}{c}{Symbol}&
\multicolumn{1}{c}{Value (Std. Error)$^{(a)}$}&
\multicolumn{1}{c}{Value (Std. Error)$^{(a)}$}\\
}
\startdata
Noise       &\quad\quad $b_0$                          &\quad  0.0000578(4)    &\quad 0.0000618(6)   \\
                &\quad\quad $b_1$                          &\quad  0.00450(8)        &\quad 0.00381(17)     \\
                &\quad\quad $b_2\,^{(b)}$                &\quad  0.170(17)         &\quad 0.0662(19)       \\
Variability$^{(c)}$ &\quad\quad $(m_{\rm s}^2)$ &\quad 0.07705(4)        &\quad 0.1045(18)       \\
Scale &\quad\quad $\kappa_0$                        &\quad  0.00566(2)        &\quad 0.00612(4)      \\
Size Parameter&\quad\quad $(k M\theta\sigma)^2$ &\quad 0.0423(2)  &\quad $\equiv 0$    \\                 
\enddata
\tablenotetext{(a)} {Standard error shown in parentheses for last digits of quoted value.}
\tablenotetext{(b)} {Includes effects of intrinsic variability on timescales between accumulation time of $512\ \mu{\rm sec}$ and integration time of 2\ sec.}
\tablenotetext{(c)} {Includes effects of intrinsic variability on timescales longer than integration time of 2\ sec.}
\label{tab:sample_parameters}
\end{deluxetable}

\clearpage

\end{document}